\documentclass[a4paper,11pt]{article}
\pdfoutput=1

\usepackage{jcappub}

\usepackage[T1]{fontenc}
\usepackage{booktabs}
\usepackage{subcaption}
\usepackage{todonotes} 
\usepackage{lipsum}
\usepackage{xspace}
\usepackage{siunitx}
\usepackage[export]{adjustbox}% http://ctan.org/pkg/adjustbox
\usepackage{textgreek}

\title{\boldmath Redshift requirements for cosmic shear with intrinsic alignment}

\newcommand{\pycosmo}{\texttt{PyCosmo}\xspace}
\newcommand{\chaoshammer}{\textsc{ChaosHammer}\xspace}
\author[a]{Silvan Fischbacher,}
\author[a,b]{Tomasz Kacprzak,}
\author[c]{Jonathan Blazek,}
\author[a]{and Alexandre Refregier}

\affiliation[a]{Institute for Particle Physics and Astrophysics, ETH Zürich,\\Wolfgang-Pauli-Strasse 27, CH-8093 Zürich, Switzerland}
\affiliation[b]{Swiss Data Science Center, Paul Scherrer Institute, 5232 Villigen PSI, Switzerland}
\affiliation[c]{Physics Department, Northeastern University,\\111 Dana Research Center, Boston, MA, 02115, USA}

\emailAdd{silvanf@phys.ethz.ch}
\emailAdd{tomaszk@phys.ethz.ch}

\abstract{
Intrinsic alignment (IA) modelling and photometric redshift estimation are two of the main sources of systematic uncertainty in weak lensing surveys.
We investigate the impact of redshift errors and their interplay with different IA models.
Generally, errors on the mean~$\delta_z$ and on the width~$\sigma_z$ of the redshift bins can both lead to biases in cosmological constraints.
We find that such biases can, however, only be partially resolved by marginalizing over~$\delta_z$ and~$\sigma_z$.
For Stage-III surveys,~$\delta_z$ and~$\sigma_z$ cannot be well constrained due to limited statistics.
The resulting biases are thus sensitive to prior volume effects.
For Stage-IV surveys, we observe that marginalizing over the redshift parameters has an impact and reduces the bias.
We derive requirements on the uncertainty of~$\sigma_z$ and~$\delta_z$ for both Stage-III and Stage-IV surveys.
We assume that the redshift systematic errors on $S_8$ should be less than half of the statistical errors, and the median bias should be smaller than~$0.25\sigma$.
We find that the uncertainty on~$\delta_z$ has to be~$\lesssim0.025$ for the NLA IA model with a Stage-III survey.
We find no requirement threshold for~$\sigma_z$ since the requirements are met even for our maximum prior width of 0.3.
For the TATT IA model, the uncertainty on~$\delta_z$ has to be~$\lesssim0.02$ and the uncertainty on~$\sigma_z$ has to be $\lesssim0.2$.
Current redshift precision of Stage-III surveys is therefore high enough to meet these requirements.
For Stage-IV surveys, systematic effects will be more important due to the higher statistical precision.
In this case, the uncertainty on~$\delta_z$ has to be~$\lesssim0.005$ and the uncertainty on~$\sigma_z$ should be $\lesssim0.1$, with no significant dependence on the IA model.
This required high precision will be a challenge for the redshift calibration of these future surveys.
Finally, we investigate whether the interplay between redshift systematics and IA modelling can explain the $S_8$-tension between cosmic shear results and CMB measurements.
We find that this is unlikely to explain the current $S_8$-tension.
The code that was used to conduct this analysis is publicly available.\footnote{\texttt{refrigerator}: \url{https://cosmo-gitlab.phys.ethz.ch/cosmo_public/refrigerator}}
}

\notoc 
\begin{document}
\maketitle
\flushbottom

\newpage
\section{Introduction}
\label{sec:intro}

Images of distant galaxies are distorted due to weak gravitational lensing by large-scale structure.
By measuring the shape correlation of millions of galaxies, the cosmic shear analysis enables us to constrain cosmological parameters (see e.g.\ \cite{bartelmann_weak_2001, refregier_weak_2003, hoekstra_weak_2008, munshi_cosmology_2008, bartelmann_gravitational_2010, kilbinger_cosmology_2015} for reviews).
Over the last years such cosmic shear surveys have become one of the most powerful probes to constrain and test cosmological models.
The main cosmological parameter of the standard $\Lambda$CDM model that cosmic shear analyses constrain is $S_8=\sigma_8 \sqrt{\Omega_m/0.3}$, where $\sigma_8$ is the amplitude of matter fluctuations on a scale of $\SI{8}{h^{-1} Mpc}$ and $\Omega_m$ is the matter density.

\begin{figure}
	\centering
		\includegraphics[width=1\textwidth]{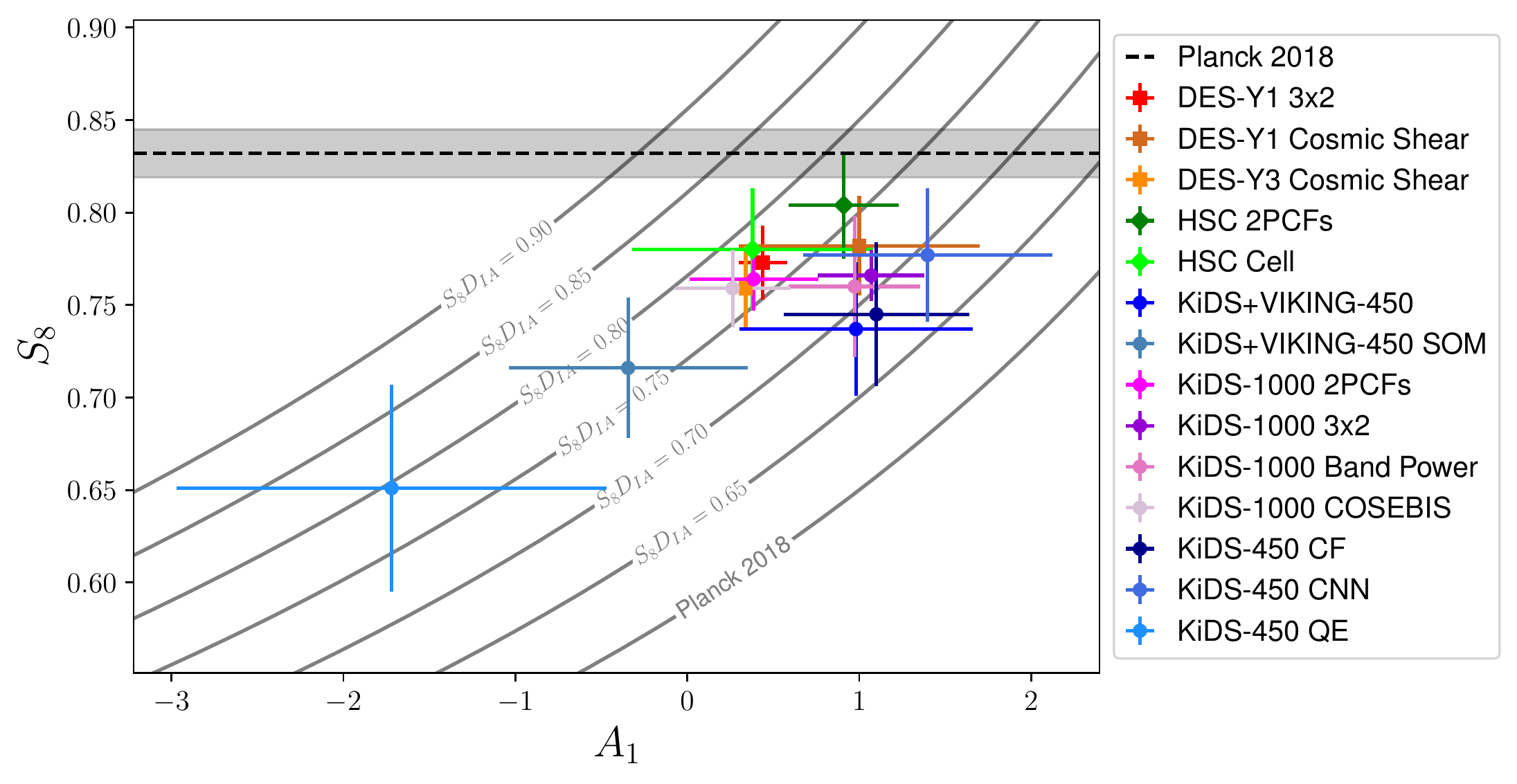}
	\caption{
	Results of different surveys in the $A_1-S_8$ plane.
	The blacked dashed with the corresponding uncertainty band corresponds to the results from Planck TT,TE,EE+lowE+lensing \cite{planck_collaboration_planck_2020}.
	The grey lines are values of constant $S_8 D_IA$ with $D_IA = 1-0.11(A_1-1)$ (see \cite{kacprzak_monte_2020}).
	The different data points are from \cite{abbott_dark_2018,troxel_dark_2018,amon_dark_nodate,secco_dark_nodate,hamana_cosmological_2020,hikage_cosmology_2019,hildebrandt_kidsviking-450_2020,wright_kidsviking-450_2020,asgari_kids-1000_2021,heymans_kids-1000_2021,hildebrandt_kids-450_2017,fluri_cosmological_2019,kohlinger_kids-450_2017}.
	For an easy comparison, we show results that use the NLA model without redshift power law to account for intrinsic alignment.
	For DES-Y1 Cosmic Shear, we plot the value with redshift power law since they do not report the exact values of the NLA result without power law.
	Note however, that without redshift power law, they find a similar $S_8$ but slightly lower $A_1$ value (see Fig.~15 and~16 of \cite{troxel_dark_2018}) which would decrease the distance to the other DES data points.
	}
	\label{fig:lensing_results}
\end{figure}

Results from recent weak lensing surveys are shown in Figure \ref{fig:lensing_results} in comparison to the CMB measurement by the Planck Collaboration \cite{planck_collaboration_planck_2020}.
It can be seen that the cosmic shear analyses find generally lower values for $S_8$ than Planck.
The Planck baseline analysis (TT+TE+EE+lowE+lensing) finds $0.832\pm 0.013$ \cite{planck_collaboration_planck_2020} while the latest cosmic shear surveys find $S_8=0.759^{+0.024}_{-0.021}$ (KiDS-1000, \cite{asgari_kids-1000_2021}) or $S_8=0.772^{+0.018}_{-0.017}$ (DES-Y3, \cite{amon_dark_nodate,secco_dark_nodate}).
This difference in the $S_8$ measurement between CMB and cosmic shear is commonly referred to as the $S_8$-tension.
Quantifying the $S_8$-tension has become an important part of modern cosmic shear analyses.
The tension can be quantified on the $S_8$ projection alone or on the full shared parameter space (see \cite{lemos2021tensions} for an overview of different tension metrics).
KiDS-1000 reports a $3\sigma$ tension with Planck comparing only the two $S_8$ 1D-projections \cite{asgari_kids-1000_2021}, DES-Y3 finds a $2.3\sigma$ tension using the full shared parameter space \cite{amon_dark_nodate, secco_dark_nodate}.

Various solutions to the $S_8$-tension have been proposed, ranging from systematics to modelling errors to new physics (for an overview, see e.g.\ \cite{leauthaud_lensing_2017, Di_Valentino_2021} and references therein).
Systematics in the photometric redshift calibration or the shear measurement are unlikely to fully explain the $S_8$-tension \cite{Hildebrandt:2020,Myles_2021, Mandelbaum_2018,Kannawadi_2019,MacCrann_2021}.
In \cite{Amon_2022} the authors argue that the tension could be solved by suppressing the nonlinear part of the matter power spectrum, e.g., by a new property of dark matter or errors in the baryonic feedback modeling.
However, the source of the $S_8$-tension is still unknown.

One of the main sources of systematic uncertainty in cosmic shear analyses is the intrinsic alignment (IA) of galaxies (see \cite{joachimi_galaxy_2015,kiessling_galaxy_2015,kirk_galaxy_2015} for reviews).
The shear measurements consist not only of the cosmic shear signal, which arise from gravitational lensing, but also of the IA signal, which is due to local physical phenomena which affect galaxy shapes.
The total power spectrum of galaxy shapes $\gamma\gamma$ is therefore a sum of lensing and IA components \mbox{$\gamma \gamma = GG + II + GI$}, 
where GG corresponds to gravitational lensing, II to the intrinsic-intrinsic galaxy alignment, and IG to the intrinsic-shear alignment.
Distinguishing lensing and IA signal is therefore one of the major challenges of current and upcoming weak lensing surveys.
Recent constraints on the IA amplitude parameter $A_1$ are also shown in Figure \ref{fig:lensing_results}.

Another source of systematic uncertainty is in the redshift estimation.
Since spectroscopic measurements of millions of galaxies are too expensive in practice, one has to infer the redshift distribution using photometric redshifts (see \cite{salvato_many_2019} for a review).
Assuming a wrong redshift distribution can lead to significantly different results, especially for the IA constraints.
This can be seen when comparing two different analyses of the KiDS+VIKING-450 data with different redshift estimation techniques.
Moving from weighted direct calibration (DIR) \cite{wright_kidsviking-450_2019} to self organizing maps (SOM) \cite{wright_kidsviking-450_2020}, the constraint on $A_1$ moves by about $2\sigma$ (see Figure \ref{fig:lensing_results}). 
By comparing the DIR and SOM redshift bins, we found that the difference in the width of these bins is between 10 and 20\%, with SOM using only about 80\% of the galaxies that are used by DIR.
The treatment of the systematics mentioned above is already a crucial part of the analysis of current Stage-III surveys such as DES \cite{amon_dark_nodate, secco_dark_nodate}, the Kilo-Degree Survey (KiDS)~\cite{asgari_kids-1000_2021} or the Subaru Hyper Suprime-Cam (HSC) \cite{hikage_cosmology_2019}.
But it will become even more important with the increased precision of upcoming Stage-IV surveys such as the Rubin Observatory's Legacy Survey of Space and Time (LSST) \cite{the_lsst_dark_energy_science_collaboration_lsst_2018}, the 
Nancy Grace Roman Space Telescope (NGRST, formerly WFIRST) \cite{spergel_wide-field_2015} or Euclid \cite{laureijs_euclid_2011}.

\begin{figure}[t!]
	\centering
    \begin{subfigure}[b]{0.32\textwidth}
         \centering
         \includegraphics[width=\textwidth]{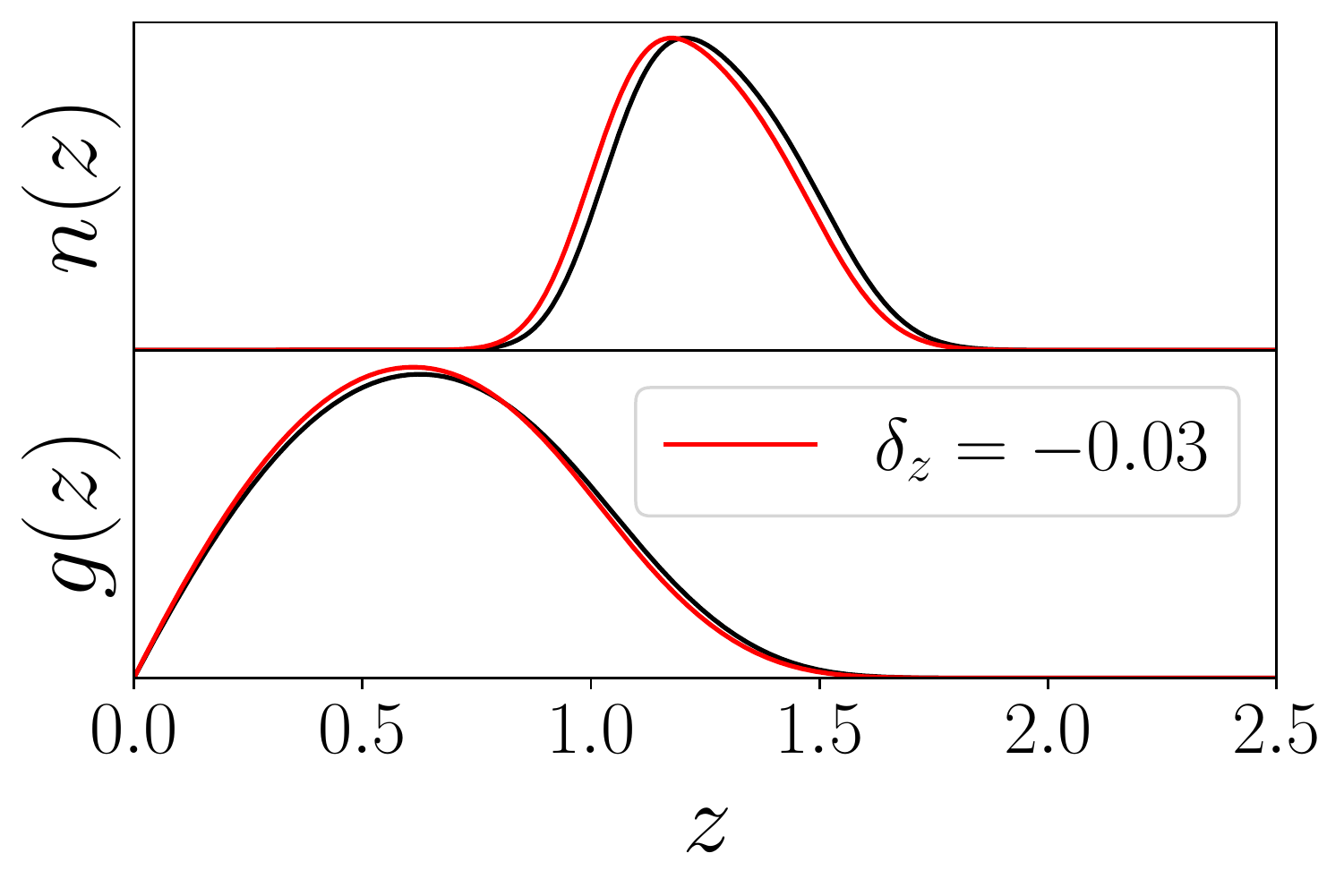}
         \caption{}
    \end{subfigure}
    \begin{subfigure}[b]{0.32\textwidth}
         \centering
         \includegraphics[width=\textwidth]{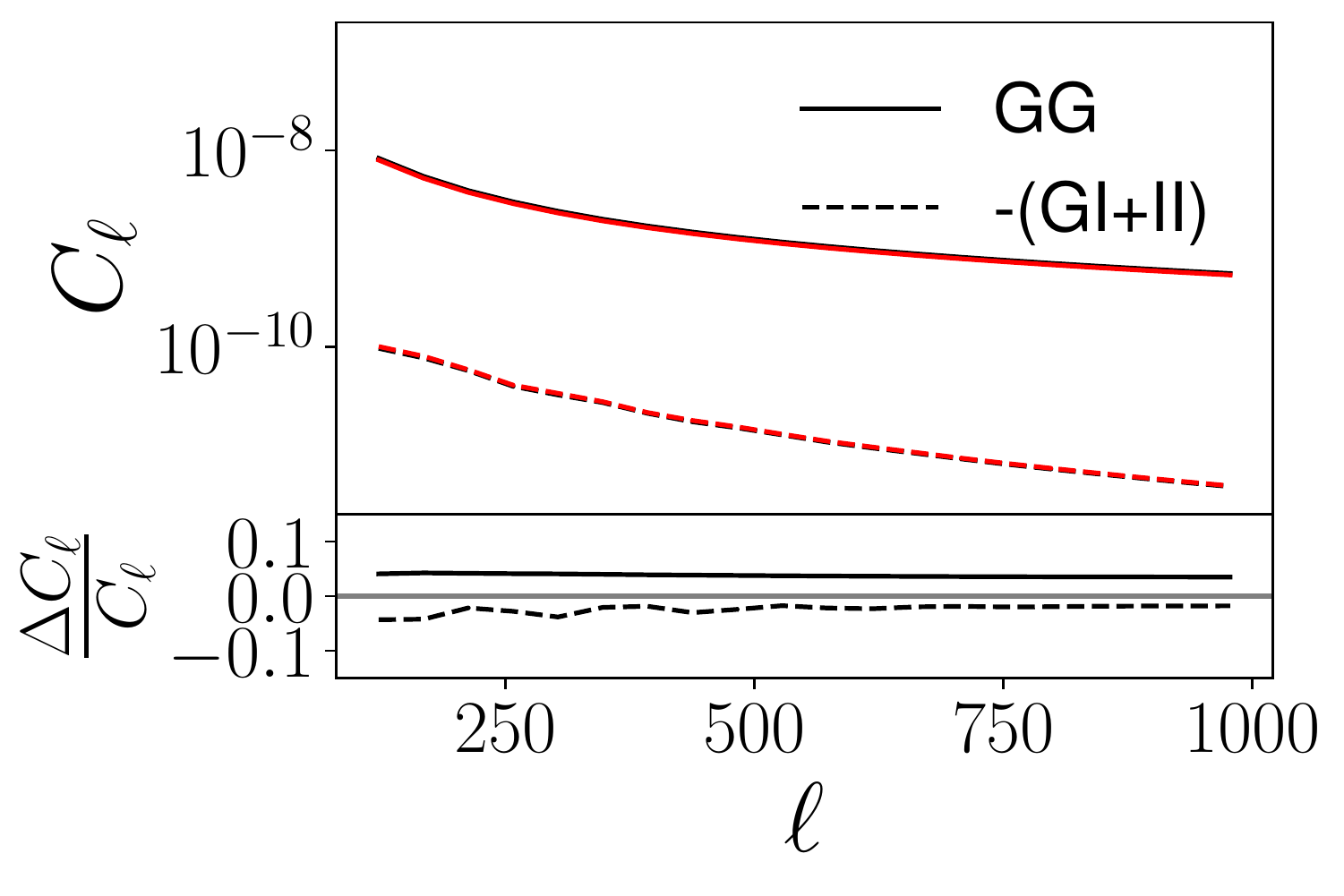}
         \caption{}
    \end{subfigure}
    \begin{subfigure}[b]{0.32\textwidth}
         \centering
         \includegraphics[width=\textwidth]{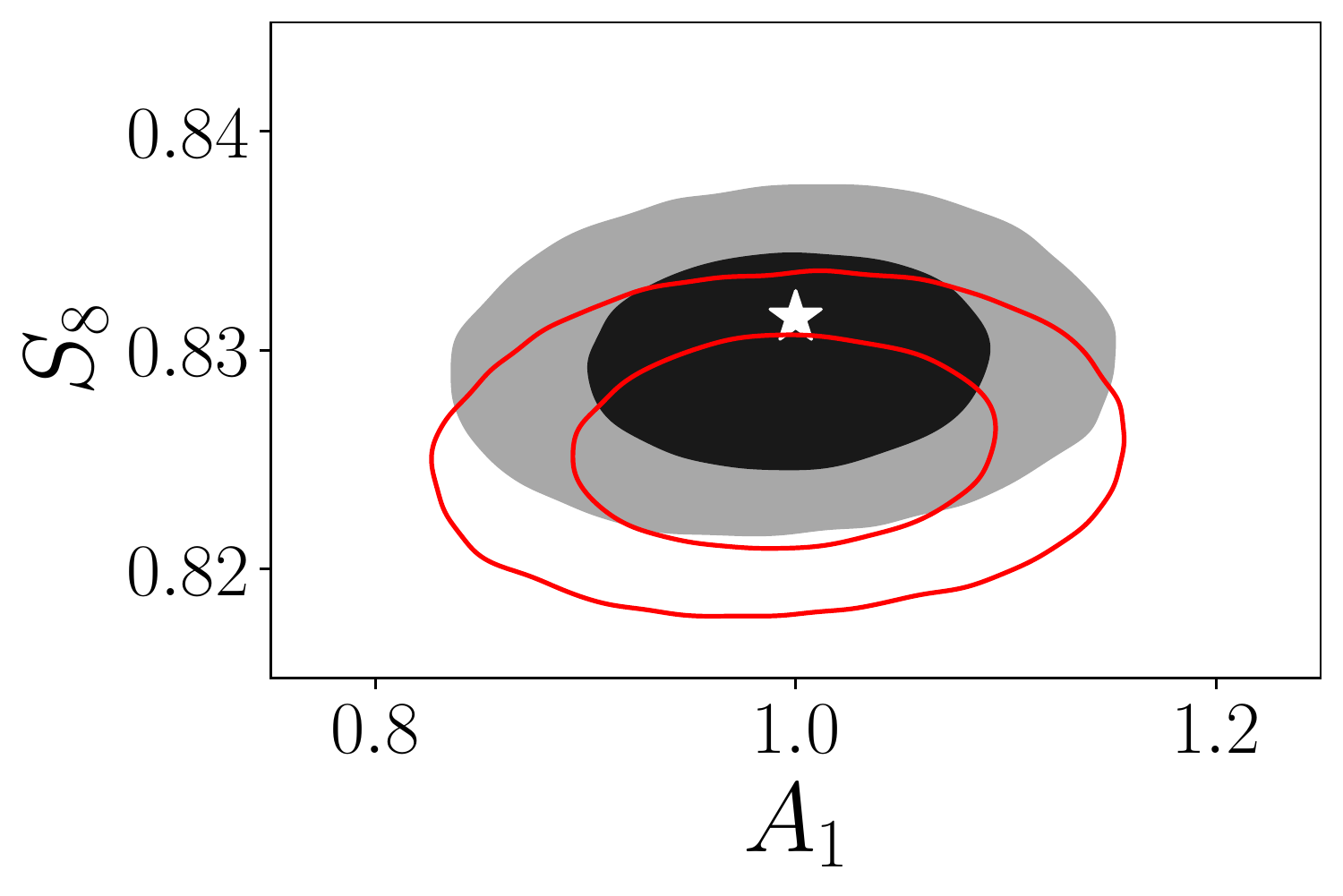}
         \caption{}
    \end{subfigure}
    \par\bigskip

    \begin{subfigure}[b]{0.32\textwidth}
         \centering
         \includegraphics[width=\textwidth]{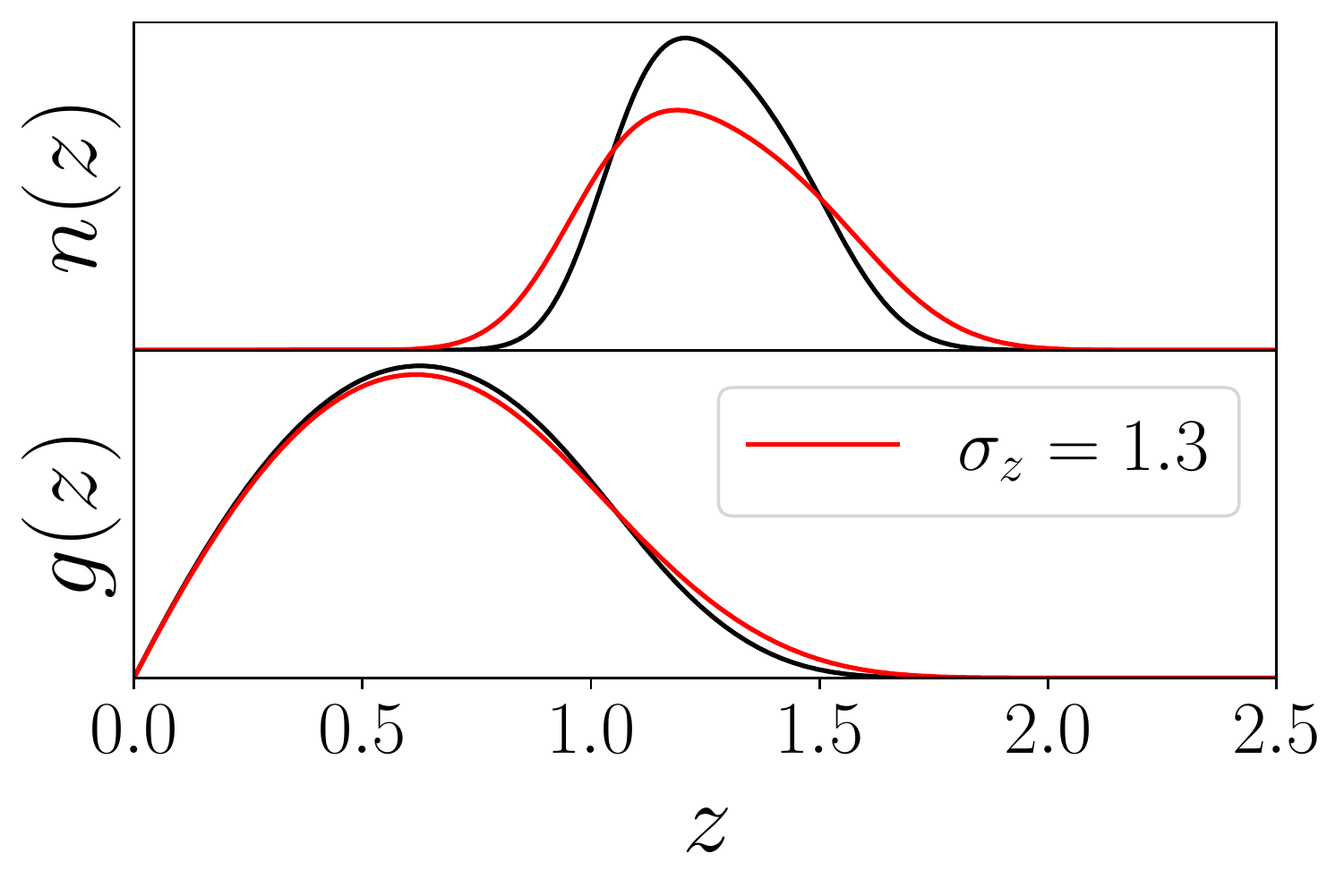}
         \caption{}
    \end{subfigure}
    \begin{subfigure}[b]{0.32\textwidth}
         \centering
         \includegraphics[width=\textwidth]{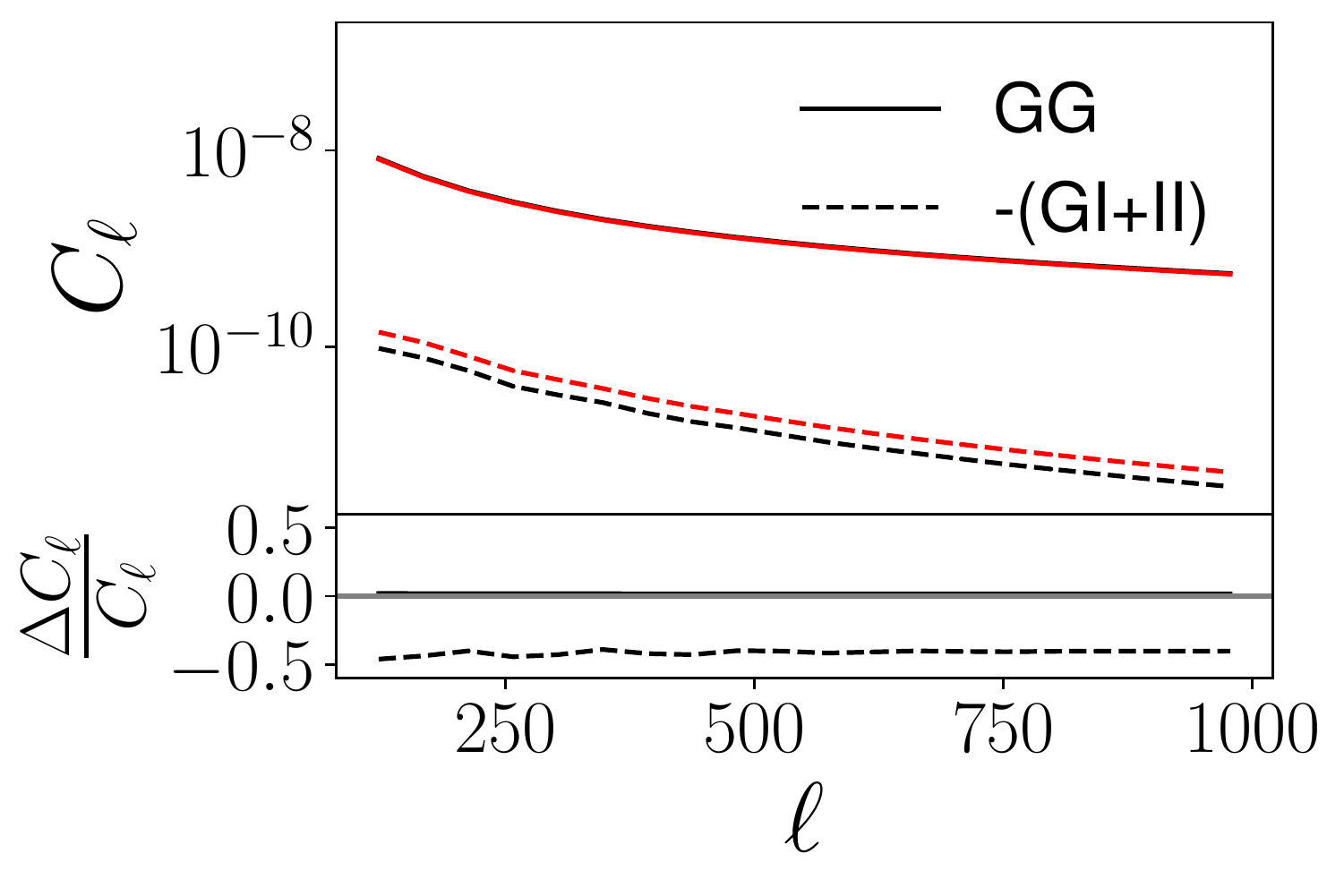}
         \caption{}
    \end{subfigure}
    \begin{subfigure}[b]{0.32\textwidth}
         \centering
         \includegraphics[width=\textwidth]{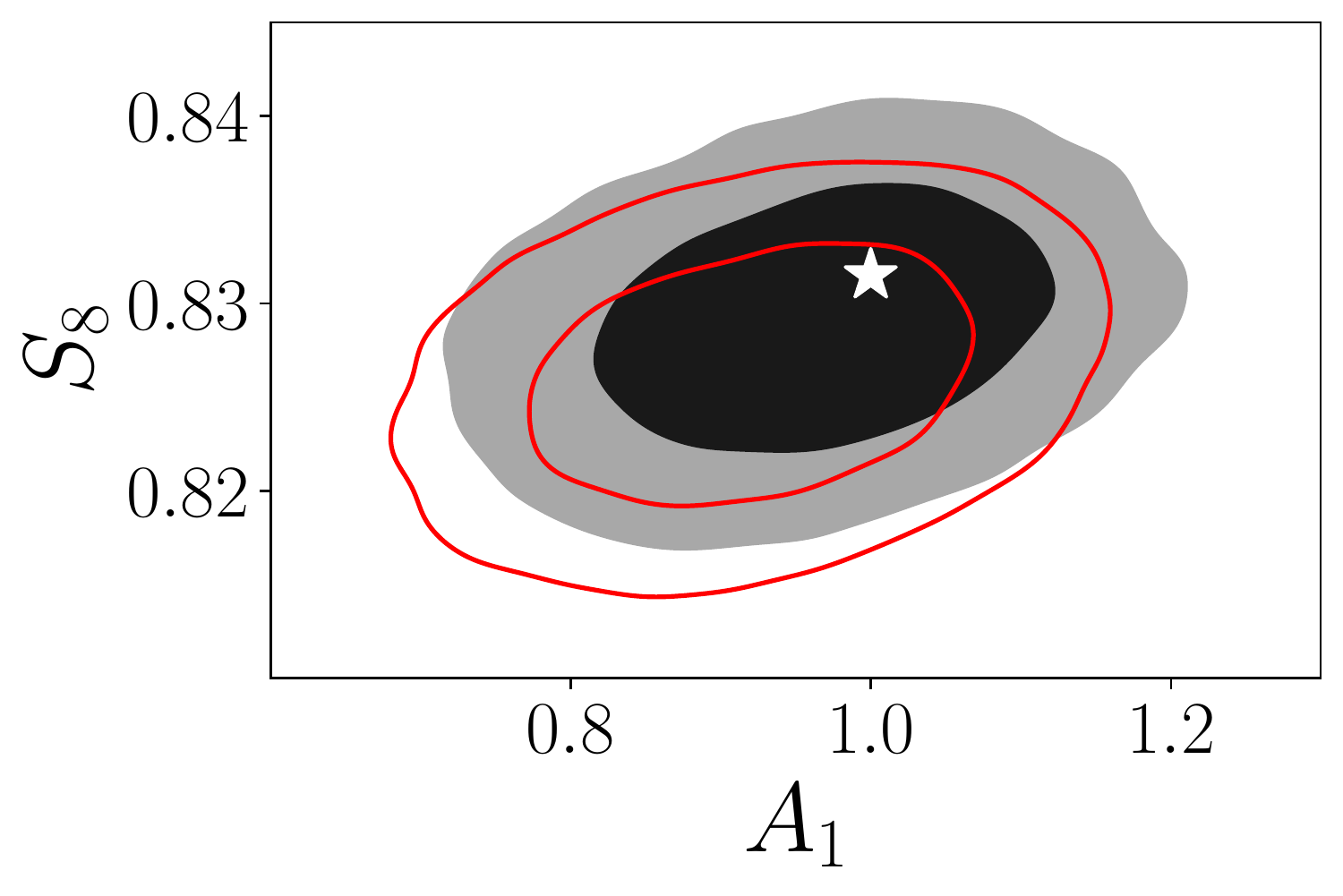}
         \caption{}
    \end{subfigure}
	\caption{
	Illustration of the impact of the error on the reshift bin mean (upper panels) and width (lower panels) on the lensing kernels, the predicted power spectra and the cosmological constraints.
	Changing the mean $n(z)$ of the mock observation, indicated by the red curve in (a), leads to a similar change in the $g(z)$ kernel (see Equation \ref{eq:lensingkernel} for the definition).
	Therefore, the difference of GG and IA component of the angular power spectrum shifts shift similarly strong, as shown in panel~(b).
	Panel~(c) shows the corresponding constraints.
	An 30\% error in the $n(z)$ width of the mock observation, as shown in panel~(d), leads only to a slight change in the kernel $g(z)$.
	The GG component of $C_{\ell}$ shifts only slightly (see (e) top) whereas the IA component shows a clearer deviation (see (e) bottom).
	The resulting constraints (68\% and 95\% confidence levels) are biased in both $S_8$ and $A_1$, as shown in panel~(f).
	This illustration is done with the fourth redshift bin of the Stage-IV setup.
	}
	\label{fig:concept}
\end{figure}

In this work, we will investigate how errors on the redshift estimation impact cosmological constraints and how it is coupled to intrinsic alignment.
We focus on cosmic shear tomography with different redshift bins and use the shear angular power spectrum $C_\ell$ for our inference.
We do not consider probe combinations of cross power spectra between galaxy shapes and galaxy positions.
An illustration of the impact of redshift errors on cosmic shear constraints is given in Figure \ref{fig:concept}, with the error on the \emph{mean}  redshift in the top panels, and errors on redshift \emph{width} in the bottom panel.

Errors on redshift bin \emph{mean} will change the predicted lensing and IA power spectra as shown in Figure \ref{fig:concept} in the upper panel.
Here, both integration kernels, and therefore the lensing and IA power spectra, change by a similar amount.
Note that lensing only integrates over the lensing kernel $g(z)$ whereas the IA spectrum integrates over $n(z)$ and $g(z)$ (see Section \ref{sec:theory} for more details).
Such errors can finally bias cosmological parameters as shown in the right panel; $S_8$ is significantly biased, while the IA amplitude $A_1$ is not affected.
Errors of the mean redshift are typically already considered in recent cosmic shear analyses (e.g.\ \cite{amon_dark_nodate, secco_dark_nodate, asgari_kids-1000_2021, hikage_cosmology_2019}) and their impact is studied (e.g.\ \cite{bonnett_redshift_2016, samuroff_simultaneous_2017}).

For errors on redshift bin \emph{width}, we find a much higher deviation of the IA power spectrum, while the GG component remains almost unchanged.
Figure~\ref{fig:concept} shows the original $n(z)$ (black line) and a $n(z)$ containing a 30\% width error (red line), as well as the corresponding lensing kernels, the GG and IA power spectra, and the resulting constraints.
Even though the GG power spectrum remains almost unchanged under the $n(z)$ error, the errors in IA power spectra cause the final $S_8$ measurement to be biased due to the wrong IA measurement.

The error in the width of $n(z)$ can thus significantly bias the measurement of $S_8$ through the coupling with intrinsic alignment. 
The goal of this paper is to quantify how both redshift errors -- mean and width -- impact cosmological constraints for different intrinsic alignment models with varying complexity.
Assuming the correct IA model, and therefore no model bias, we compute requirements on the $n(z)$ mean and width calibration for the current and upcoming surveys.
We consider four commonly used IA cases: no IA, nonlinear alignment model (NLA) \cite{bridle_dark_2007}, NLA with redshift evolution (NLA-z), and the Tidal Alignment and Tidal Torque (TATT) model \cite{blazek_beyond_2019} (see Section~\ref{sec:theory} for descriptions of these models).
We also investigate if such redshift errors and their coupling with IA could explain the current $S_8$ tension.

The paper is organized as follows.
In Section~\ref{sec:theory}, we give a short overview of the theory behind cosmic shear and present the different intrinsic alignment models used in this work.
We present our methodological approach in Section~\ref{sec:methods} where we explain the different parts of the pipeline including angular power spectrum prediction, redshift errors, covariance matrix estimation and the inference process.
Our results are shown in Section~\ref{sec:response}-\ref{sec:lowS8}.
First, we show how redshift errors on individual bins impact constraints of $S_8$ and $A_1$ in Section~\ref{sec:response}.
We show how ignoring $n(z)$ shape errors can bias cosmological constraints in Section~\ref{sec:ignore}.
In Section~\ref{sec:include}, we investigate how biases from Section~\ref{sec:ignore} change if we add redshift width errors in the modelling process.
We compute requirements on redshift estimation for current and future surveys in Section~\ref{sec:req}.
Finally, we explore if an interplay between IA modelling and redshift errors could be the reason for the $S_8$-tension in Section~\ref{sec:lowS8} before we conclude in Section~\ref{sec:conclusions}.

\section{Theory}
\label{sec:theory}
We use the angular cosmic shear power spectrum $C_\mathrm{GG}(\ell)$ in a tomographic setup all through our analysis. The auto- and cross correlation between bin $i$ and $j$ can then be expressed in terms of the three-dimensional matter power spectrum $P(k,\chi)$ by
\begin{equation}\label{eq:Cell}
C_\mathrm{GG}^{i,j}(\ell)= \int_{0}^{\chi_\mathrm{hor}} d \chi\frac{g^i(\chi)g^j(\chi)}{\chi^2} P\left(\frac{\ell}{\chi}, \chi\right),
\end{equation}
where $\chi_\mathrm{hor}$ is the comoving distance to the horizon and $g(\chi)$ the lensing kernel given by
\begin{equation}\label{eq:lensingkernel}
g(\chi)=\frac{3}{2} \frac{H_{0}^{2} \Omega_{\mathrm{m}}}{c^{2}} \frac{\chi}{a(\chi)} \int_{\chi}^{\chi_{\mathrm{hor}}} d \chi^{\prime} n\left(\chi^{\prime}\right) \frac{\chi^{\prime}-\chi}{\chi^{\prime}}, 
\end{equation}
where $H_0$ is the Hubble parameter today, $c$ the speed of light, $a$ is the scale factor, $\Omega_m$ the matter density parameter and $n(\chi)$ the redshift distribution of the sample. We assume a flat $\Lambda$CDM Universe.

\subsection{Intrinsic alignment models}
\label{sec:IAmodels}
In a homogeneous and isotropic Universe, we would expect that galaxies are randomly oriented and the noise of the orientation and shape of individual galaxies would compensate for high statistics. But local physical phenomena tend to align galaxies on small scale. This is called intrinsic alignment (see e.g.\ \cite{joachimi_galaxy_2015,kiessling_galaxy_2015,kirk_galaxy_2015} for reviews).
The observed angular power spectrum is then a sum of the auto- and cross-correlations of the cosmic shear and the intrinsic ellipticity of the source given by
\begin{equation}\label{eq:ellipticity}
C_\ell^\mathrm{obs} = C_\mathrm{GG}^{i,j}(\ell) + C_\mathrm{GI}^{i,j}(\ell) + C_\mathrm{IG}^{i,j}(\ell) + C_\mathrm{II}^{i,j}(\ell)
\end{equation}
where $C_\mathrm{GG}^{i,j}$ corresponds to the pure cosmic shear signal and the II and GI component are typically expressed as
\begin{align}\label{eq:II}
    C_{\mathrm{II}}^{i,j}(\ell) &= \int_{0}^{\chi_\mathrm{hor}} d \chi\frac{n^i(\chi)n^j(\chi)}{\chi^2} P_\mathrm{II}\left(\frac{\ell}{\chi}, \chi\right),\\
\label{eq:GI}
    C_{\mathrm{GI}}^{i,j}(\ell) &= \int_{0}^{\chi_\mathrm{hor}} d \chi\frac{g^i(\chi)n^j(\chi)}{\chi^2} P_\mathrm{GI}\left(\frac{\ell}{\chi}, \chi\right).
\end{align}
where $g^i(\chi)$ is the lensing kernel and $n^i(\chi)$ the galaxy distribution.
We use different models to describe the power spectra $P_\mathrm{II}$ and $P_\mathrm{GI}$ that we will present in the next paragraphs.
A summary of the models and their free parameters is given in Table \ref{tab:IAmodels}.
All the models that we consider in this work have been successfully applied to real data in the last years.

\begin{table}[t!]
    \centering
    \begin{tabular}{ll}
    \toprule
    \textbf{Model Name} & \textbf{Free IA Parameters} \\
    \midrule
    noIA & - \\ 
    NLA  & $A_1$ \\
    NLA-z & $A_1,\eta_1$\\
    TATT & $A_1, A_2, \eta_1, \eta_2$ and $b_\mathrm{ta}$\\
    \bottomrule
    \end{tabular}
    \caption{Summary of all intrinsic alignment models used in this work.}
    \label{tab:IAmodels}
\end{table}

\subsubsection{Nonlinear alignment model (NLA)}
The nonlinear alignment model (NLA) \cite{bridle_dark_2007} expresses the intrinsic alignment power spectra in terms of the matter power spectrum by
\begin{align}
    P_\mathrm{GI}(k,z) &= A(z) P(k,z),\\
    P_\mathrm{II}(k,z) &= A^2(z) P(k,z).
\end{align}
The standard NLA model uses for the normalization the following convention
\begin{equation}\label{eq:A1}
A(z) = - A_1 C_1 \frac{\rho_\mathrm{crit}\Omega_m}{D(z)},
\end{equation}
where $A_1$ is the intrinsic alignment amplitude and a free parameter in the analysis, $C_1=5\times 10^{-14} M_{\odot}^{-1} h^{-2}\, \si{Mpc}$ is fixed to a value from \cite{brown_measurement_2002}, $\rho_\mathrm{crit}$ is the critical density of the Universe and $D(z)$ is the linear growth factor.

We use an extended NLA model, in the following called NLA-z, to account for possible redshift dependence of the intrinsic alignment signal. The normalization is then given by
\begin{equation}\label{eq:A1z}
A(z) = - A_1 C_1 \frac{\rho_\mathrm{crit}\Omega_m}{D(z)} \left ( \frac{1+z}{1+z_0} \right)^{\eta_1},
\end{equation}
where $\eta_1$ is an additional free parameter and the pivot redshift $z_0$ is usually fixed at a parameter corresponding to the mean redshift of a survey. To allow easy comparison, we fix $z_0=0.62$ for all our setups following the choice of the DES survey \cite{troxel_dark_2018}.

The NLA model is widely used to account for the IA signal, e.g.\ in the KiDS-1000 \cite{asgari_kids-1000_2021} or HSC-Y1 \cite{hikage_cosmology_2019,hamana_cosmological_2020} analysis.
However, it only accounts for linear alignment.
This describes well the alignment of red galaxies at large scales \cite{joachimi_galaxy_2015} and since the IA signal is much stronger for red than for blue galaxies \cite{Hirata:2007}, the NLA model is a sufficient model for these Stage-III surveys.

\subsubsection{Tidal alignment \& tidal torque model (TATT)}
The tidal alignment and tidal torque (TATT) model is based on a perturbative expansion to account for higher order contributions \cite{blazek_beyond_2019}.
It includes the tidal alignment that is linear in the tidal field, the tidal torque \cite{mackey_theoretical_2002,codis_spin_2015} that is quadratic in the tidal field, and the impact of source density weighting \cite{blazek_tidal_2015}.
The TATT model has three normalization parameters given by
\begin{align}\label{eq:A1_TATT}
&A_{1}(z)=-A_{1} {C}_{1} \frac{\rho_{\text {crit }} \Omega_{\mathrm{m}}}{D(z)}\left(\frac{1+z}{1+z_{0}}\right)^{\eta_{1}}, \\
\label{eq:A2_TATT}
&A_{2}(z)=5 A_{2} {C}_{1} \frac{\rho_{\text {crit }} \Omega_{\mathrm{m}}}{D(z)^{2}}\left(\frac{1+z}{1+z_{0}}\right)^{\eta_{2}}, \\
\label{eq:A1d_TATT}
&A_{1 \delta}(z)=b_{\mathrm{ta}} A_{1}(z), 
\end{align}
where $A_1, A_2, \eta_1, \eta_2$ and $b_\mathrm{ta}$ are free parameters in our analysis and $z_0=0.62$ is fixed as for NLA-z. Note that the TATT model reduces to the NLA model for $A_2=\eta_2=b_\mathrm{ta}=0$.

The TATT model is a natural extension of the NLA model, which only accounts for tidal alignment.
It can describe tidal torque effects which are expected for blue galaxies \cite{joachimi_galaxy_2015} as well as the impact of source density weighting and related alignment mechanisms \cite{blazek_tidal_2015}.
It has already been successfully applied to real data in the DES-Y3 analysis.
The authors showed that using the simpler NLA model could bias cosmological results significantly if the IA signal follows the mean IA posterior of DES-Y1 3x2pt analysis by \cite{samuroff_2019}.
However, after unblinding the Y3 data, they found that using the NLA model would not have biased the final result since the IA amplitude was lower than expected \cite{secco_dark_nodate}.
For future surveys with increasing precision, more complex IA models like the TATT model will likely be required to ensure that the results are not biased.
There are also other models developed that could be used in future analyses, e.g.\ the halo model \cite{Fortuna:2020} that especially accounts for small scale contributions by satellite galaxies or the effective field theory (EFT) approach described in \cite{Vlah:2019}.
The EFT model is similarly constructed as the TATT model, but it is more general and captures all possible contributions to IA at next-to-leading order, including those contained in the TATT model. 

\subsection{Probabilistic problem formulation and requirements}
\label{sec:probabilistic_formulation}

In the Bayesian inference framework, we obtain the posterior on the \mbox{parameters of interest $\theta$} by multiplying the likelihood of the observed data by the prior on $\theta$. 
The observed data is typically affected by two components: signal and noise.
The noise is composed of the cosmic variance and the observational effects, such as galaxy shape noise and measurement noise.
By splitting the data into signal and noise, we can express the inference problem as:
\begin{equation}
    p(\theta \ | \ C_\ell^{\rm{signal}}, C_\ell^{\rm{noise}}) = p(C_\ell^{\rm{signal}}, C_\ell^{\rm{noise}}| \theta) p(\theta).
\end{equation}
A given signal vector will yield a different posterior distribution for different noise realization.
Therefore, in order to be able to compute requirements for a given true cosmology fitted by a given model, we cannot rely on a single noise realization, and need to aggregate noise realizations in some way.
The simplest way to do this is to calculate an ``average posterior'', where we marginalize the noise distribution:
\begin{equation}
\label{eqn:average_posterior}
   p(\theta \ | \ C_\ell^{\rm{signal}}) = \int p(\theta \ | \ C_\ell^{\rm{signal}}, C_\ell^{\rm{noise}}) \ dC_\ell^{\rm{noise}}  = \int p(C_\ell^{\rm{signal}}, C_\ell^{\rm{noise}}| \theta) p(\theta) \ dC_\ell^{\rm{noise}},
\end{equation}
which can be easily calculated by obtaining MCMC chains for multiple realizations of $C_\ell^{\rm{noise}}$, and concatenating the samples.
To obtain the requirements, we will calculate parameters biases using the average posterior in Equation~\ref{eqn:average_posterior}.
The average posterior will be larger than the posterior from a single observation, which determines the expected uncertainty from a single survey.
 
\section{Method}
\label{sec:methods}

\subsection{Theoretical predictions with \pycosmo}
\label{sec:pycosmo_implementation}
The theoretical predictions of the angular power spectrum $C_\ell$ are computed using \pycosmo \cite{refregier_pycosmo_2018,tarsitano_predicting_2020,moser_symbolic_2021} using the Limber approximation \cite{limber_analysis_1953,kaiser_weak_1992,kaiser_weak_1998}.
The linear power spectrum is computed using the Eisenstein \& Hu model \cite{eisenstein_power_1999} for the transfer function, for the nonlinear power spectrum we use the revised halofit model \cite{takahashi_revising_2012}.
We bin the $C_\ell$ according to the binning scheme presented in Appendix \ref{app:setup}.
For the implementation of the TATT model, we use \texttt{FAST-PT} \cite{mcewen_fast-pt_2016,fang_fast-pt_2017} to compute the $k$-dependent terms of Equation 37-39 in \cite{blazek_beyond_2019}.

\subsection{Redshift bins}
\label{sec:redshfit}
\begin{figure}[t!]
	\centering
% 	\begin{subfigure}[b]{0.3\textwidth}
%          \centering
%          \includegraphics[width=\textwidth]{figures/redshift_width_KiDS.pdf}
%          \caption{KiDS-1000}
%     \end{subfigure}
    \begin{subfigure}[b]{0.48\textwidth}
         \centering
         \includegraphics[width=\textwidth]{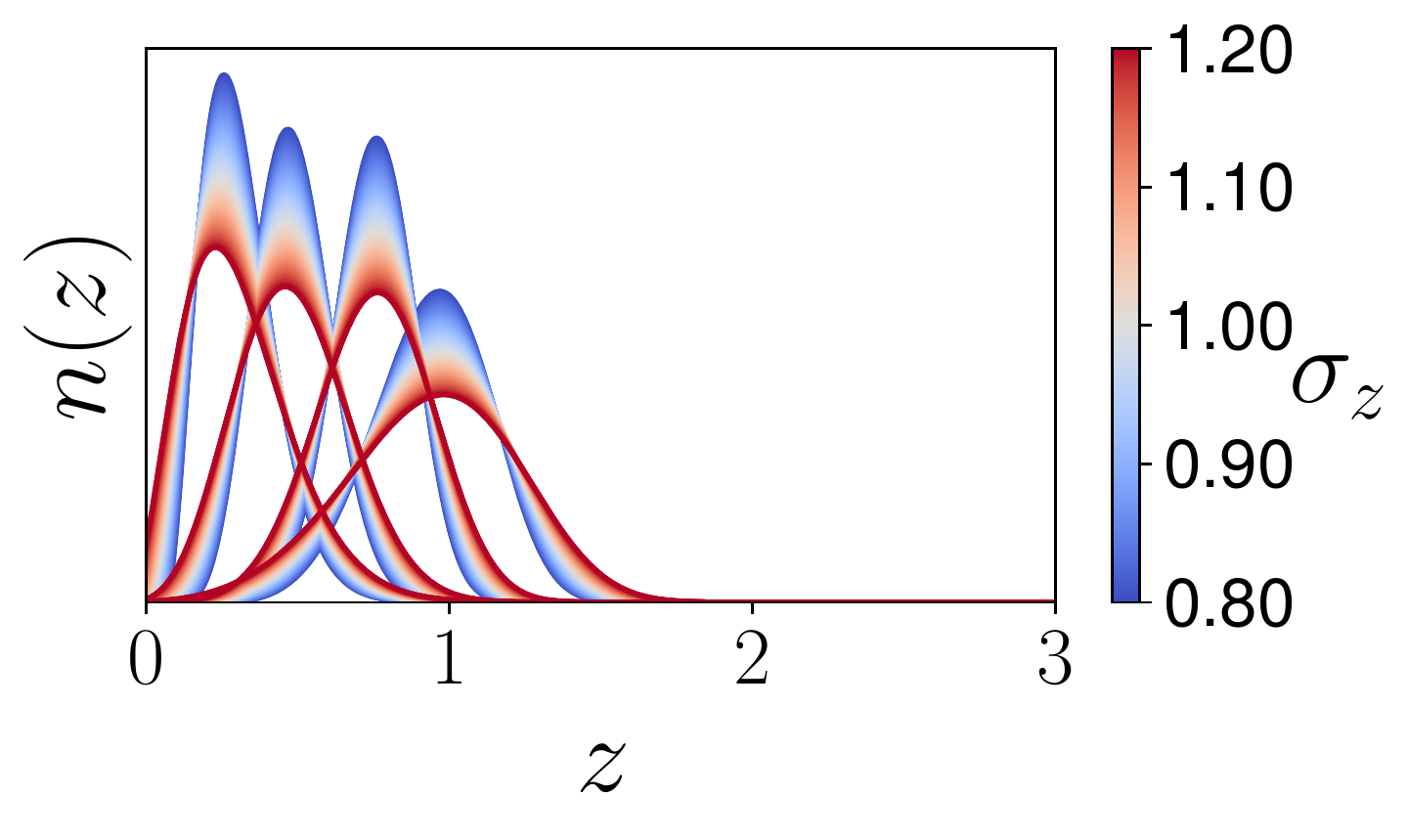}
         \caption{Stage-III}
    \end{subfigure}
    \begin{subfigure}[b]{0.48\textwidth}
         \centering
         \includegraphics[width=\textwidth]{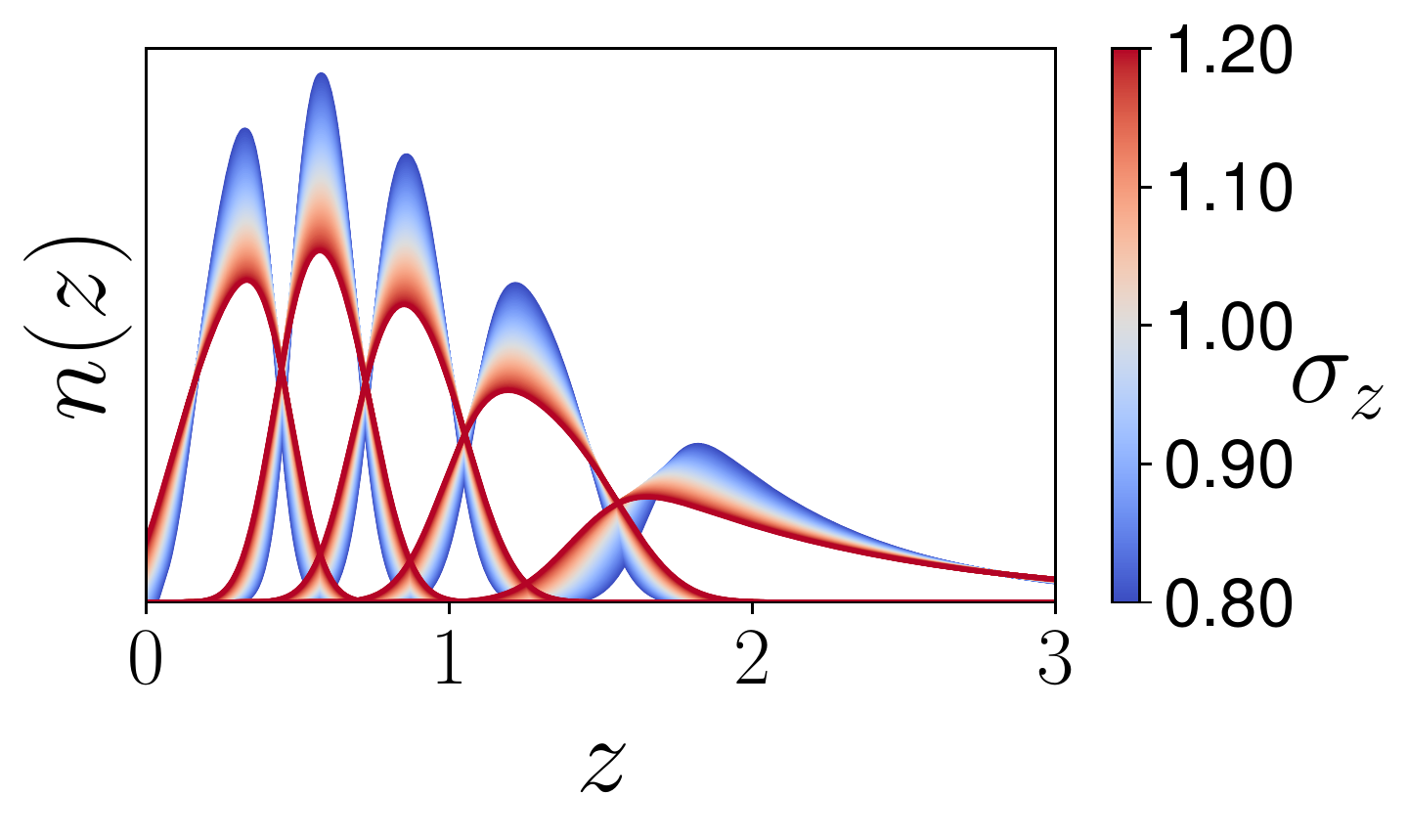}
         \caption{Stage-IV}
    \end{subfigure}
	\caption{Impact of redshift bin width parameter $\sigma_z$ on the bins used in this paper.}
	\label{fig:zsigma}
\end{figure}

We use two types of redshift bin distributions which are shown in Figure \ref{fig:zsigma}.
The redshift bins of the Stage-III setup are a smoothed version of the four DES-Y3 bins \cite{amon_dark_nodate}.
The redshift bins of the Stage-IV bins are created assuming the expected total redshift distribution of LSST~\cite{the_lsst_dark_energy_science_collaboration_lsst_2018} which are then cut in five equally populated redshift bins.
Using five bins for Stage-IV follows the analysis choices given in \cite{the_lsst_dark_energy_science_collaboration_lsst_2018} noting that other Stage-IV forecasts also use more bins (e.g.~\cite{zurcher_towards_2022} where they compare the use of 5~and 10~bins).
We refer to Appendix~\ref{app:setup} for a more detailed description.

We study two types of redshift distribution errors characterized by the parameters~$\delta_z$ and~$\sigma_z$. 
Parameter~$\delta_z$ describes the error of the mean redshift and is already part of most cosmic shear analyses.
Errors of the width of the redshift bins are typically not considered.
We define such errors by the parameter~$\sigma_z$.
It corresponds to a simple stretch of the redshift bin by a factor of $\sigma_z$ of the $z$-axis.
The standard value corresponding to no error in the redshift estimation corresponds therefore to~$\sigma_z=1$.
Since the integration range of Equation~\ref{eq:Cell} and the corresponding equations for IA do not cover negative redshifts, we cut all our input redshift bins at $z=0$.
This can have the consequence that if you stretch a bin with~$\sigma_z>1$, a large part of the bin would overlap with $z=0$ and is therefore cut.
This leads to a change in the mean of the bin to higher redshift.
That is why our stretching method involves a shift to lower redshift until the mean redshift of unstretched and stretched bin are the same.
The bin is also renormalized after the cut.
This way, we can independently control the error of the mean redshift with $\delta_z$ and the error of the width of the redshift bin with~$\sigma_z$.
The impact of the parameter $\sigma_z$ is shown in Figure~\ref{fig:zsigma}, where $\sigma_z \in [0.8,1.2]$. Other ways of varying the width of the redshift bins were tested, but this stretching method has proven to work best for our approach with an intuitive meaning of the parameter and fast computation time.

\subsection{Covariance matrix estimation}
\label{sec:covmat}
The covariance matrix is estimated from simulations which has proven to give accurate results (see e.g.\ \cite{zurcher_cosmological_2021}).
We use the lightcones from the N-Body simulation described in \cite{zurcher_cosmological_2021} generated by the N-Body code \textsc{PKDGRAV3} \cite{potter_pkdgrav3_2016}.
We generate 50~full-sky weak lensing maps for each redshift bin from these lightcones using \texttt{UFalcon} \cite{sgier_fast_2019,sgier_fast_2021}.
These full-sky maps are then cut into patches of the sky corresponding to the survey specifications; see Table~\ref{tab:covmat} for numerical values. 
The angular power spectrum $C_\ell$ is then computed for each of these patches using the \texttt{healpy} package\footnote{\url{https://healpix.sourceforge.io/}} \cite{gorski_healpix_2005,zonca_healpy_2019}.
This results in 2000~$C_\ell$ realisations that are used to estimate the covariance matrix of the cosmic variance.

The shape noise component is estimated from noise maps generated with shape noise per component $\sigma_\varepsilon$ and galaxy number density $n_\mathrm{eff}$ as given in Table \ref{tab:covmat}.
These noise maps are again cut into patches of corresponding size and the angular power spectrum is computed for each of these patches.
We use the \texttt{healpy} package for both creation and computation of the angular power spectrum.
The covariance matrix accounting for noise is then computed from 6000~such $C_\ell$ realisations.
We obtain the final covariance matrix by a simple addition of the noise and cosmic variance covariance matrix assuming that shape noise and cosmic variance are uncorrelated. 

\subsection{$C_{\ell}$ emulator}
\label{sec:C_ell_emulator}
To perform inference on multiple mock observations with different theory models, we use a fast emulator of the tomographic angular power spectrum $C_\ell$, dubbed \chaoshammer.
Similar emulators were created in the past \cite{Kwan2015emulation,Marulli2021anns,Knabenhans2021euclid2}.
This emulator is trained on $C_\ell$ predicted by \pycosmo.
It takes $\sim5$ seconds to calculate the set of auto and cross $C_\ell$ for a given cosmology.
After training the fast emulator generates the same $C_\ell$ in about 1~millisecond on a CPU for \pycosmo.
The emulator is based on a shallow neural network that maps from a set of input parameters, containing cosmology, intrinsic alignment and redshift errors, to binned $C_\ell$ values; 200~for Stage-III redshift bins and 750~for Stage-IV bins. 
The neural network was implemented in \texttt{Tensorflow} \cite{tensorflow_developers_tensorflow_2021} and trained on 2 to 100~million points depending on the number of free parameters and the width of the prior.
Those points were sampled from a Halton sequence using the package \texttt{ChaosPy} \cite{feinberg_chaospy_2015} and evaluated by \pycosmo to full convergence to predict the $C_\ell$. 
After training, the emulator reaches the accuracy of $A>98.5\%$, where the $A$ is the median absolute deviation of the fractional difference between the $C_\ell$ generated by \pycosmo and by the prediction of the emulator, $A=1-(C_\ell^{\rm{emu}}-C_\ell)/C_\ell$ calculated over the entire training set.
This fast emulator enables us to run multiple MCMC chains to full convergence very quickly, in an order of minutes for a chain with $10^6$~samples on a CPU.
The remaining emulator error of $<1.5\%$ (Stage-III) or $<0.7\%$ (Stage-IV) will not have an impact on our inference, given that for Stage-III and IV the uncertainty on the $C_\ell$ element is much larger than that. 
For details of construction and training of the \chaoshammer models, see Appendix~\ref{app:chaoshammer}.

\subsection{Cosmological inference and best-fit}
\label{sec:bestfit}
We run MCMC chains using the covariance matrix estimated as described in Section~\ref{sec:covmat} and the $C_\ell$ emulator (see Section~\ref{sec:C_ell_emulator}).
We list prior ranges and fiducial values in Table~\ref{tab:prior}.
Note that depending on the setup, parameters are either fixed to their fiducial values or are varied in the analysis according to their prior range.
We state for each setup which parameters are fixed and which are varied in the prior range (for a summary, see Table \ref{tab:prior_section}).

The best-fit is obtained by a Nelder-Mean minimization \cite{nelder_simplex_1965} of the $\chi^2$ starting from the MCMC sample with the highest $\chi^2$.
This method has proven to be very stable for a noiseless data vector and also recovers the true parameters in test setups without bias.

In setups where we the data vector contains noise, the minimization is less stable but still better than using just the best sample of the MCMC.
If very high precision of the best-fit is necessary, the $\chi^2$ minimization is repeated several times from different starting points and the mode of the resulting distribution is used as best-fit.
We choose the mode over the median to avoid complications for bimodal distributions.
This gives stable results even for a noisy data vector.

\begin{table}[t]
    \centering
    \begin{tabular}{lllll}
        \toprule
        \textbf{Parameter} & \textbf{Fiducial Value} & \textbf{Range: Stage-III} & \textbf{Range: Stage-IV} & \textbf{Prior} \\
        \midrule
	    $\Omega_m$ & 0.3153 & $[0.1,0.6]$ & $[0.15,0.45]$ & flat \\
		$S_8$ & 0.8315 & $[0.6,1.0]$ & $[0.7,0.9]$ & flat\\
		$n_s$ & 0.9649 & $[0.9,1.05]$ & $[0.92,1.0]$ & flat\\
		$h$ & 0.6736 & $[0.6,0.85]$ & $[0.65,0.75]$ &flat\\
		$\Omega_b$ & 0.0493 & $[0.03,0.07]$ & $[0.04,0.05]$ & flat\\
		\midrule
		$A_i$ &  1 & $[-4,4]$ & $[-2,2]$ & flat\\
		$\eta_i$ & 0 & $[-4,4]$ & $[-4,4]$ & flat\\
		$b_\mathrm{ta}$ & 1 & $[0,2]$ & $[0,2]$ & flat\\
		$z_0$ & 0.62 & 0.62 & 0.62 & fixed\\
		\midrule
		$\delta_{z,i}$ & 0 & $[-0.03,0.03]$ & $[-0.03,0.03]$ & $0\pm0.01$ \\
		$\sigma_{z,i}$ & 1 & $[0.5,2]$ & $[0.5,2]$ & $1\pm0.1$\\
		\bottomrule
    \end{tabular}
    \caption{Priors used in the analysis.
    If a parameter is varied or fixed depends on the actual setup where it is used.
    We list all relevant numbers for all parameters here.
    The fiducial value is used in the case when the parameter is fixed.
    Cosmological parameters are fixed to the fiducial result from Planck TT+TE+EE+lowE+lensing \cite{planck_collaboration_planck_2020}, intrinsic alignment is fixed to standard values and the redshift errors are fixed to values that corresponds to the unchanged redshift bins.
    The range defines the upper and lower limits that are used in the analysis which is different for the Stage-III and Stage-IV setup.
    The prior characterizes the distribution that is used if the parameter is varied.
    Priors of the form $m\pm s$ correspond to Gaussian priors with mean $m$ and standard deviation $s$.}
    \label{tab:prior}
\end{table}

\begin{table}
    \begin{subtable}[t]{0.48\textwidth}
    \centering
    \begin{tabular}{l|ll}
    \toprule & \textbf{mock observation} & \textbf{analysis}\\
    \hline
    cosmo. & fixed & flat \\
    IA & fixed & flat\\
    $\delta_z$ & fixed or flat & $0\pm 0.01$\\
    $\sigma_z$ & fixed or flat & fixed\\
    \bottomrule
    \end{tabular}
    \subcaption{
    Section \ref{sec:response}:
    For each subplot in Figures \ref{fig:impact_of_1zparam_NLA} and \ref{fig:impact_of_1zparam_TATT}, there is only one redshift parameter varied.
    All other parameters are fixed.
    }
    \label{tab:response}
    \end{subtable}
    \begin{subtable}[t]{0.48\textwidth}
        \centering
    \begin{tabular}{l|ll}
    \toprule & \textbf{mock observation} & \textbf{analysis}\\
    \hline
    cosmo. & fixed & flat \\
    IA & fixed & flat\\
    $\delta_z$ & fixed & $0\pm 0.01$\\
    $\sigma_z$ & fixed to Eq. \ref{eq:testsetup} & fixed\\
    \bottomrule
    \end{tabular}
    \subcaption{
    Section \ref{sec:ignore}
    }
    \label{tab:ign}
    \end{subtable}
    
    \begin{subtable}[t]{0.48\textwidth}
    \centering
    \begin{tabular}{l|ll}
    \toprule & \textbf{mock observation} & \textbf{analysis}\\
    \hline
    cosmo. & fixed & flat \\
    IA & fixed & flat\\
    $\delta_z$ & fixed & $0\pm 0.01$\\
    $\sigma_z$ & fixed to Eq. \ref{eq:testsetup} & $1\pm0.1$\\
    \bottomrule
    \end{tabular}
    \caption{
    Section \ref{sec:include}
    }
    \label{tab:incl}
    \end{subtable}
    \begin{subtable}[t]{0.48\textwidth}
    \centering
    \begin{tabular}{l|ll}
    \toprule & \textbf{mock observation} & \textbf{analysis}\\
    \hline
    cosmo. & fixed & flat \\
    IA & flat & flat\\
    $\delta_z$ & flat & $0\pm 0.01$\\
    $\sigma_z$ & flat & fixed\\
    \bottomrule
    \end{tabular}
    \caption{
    Section \ref{sec:lowS8}
    }
    \label{tab:lowS8}
    \end{subtable}

    \caption{
    Summary of priors on cosmological (cosmo.), intrinsic alignment (IA) and the two redshift parameters $\delta_z$ and $\sigma_z$ used in different sections for mock observation and analysis.
    The values for the fixed parameters are given in Table \ref{tab:prior} as well as the range of the flat priors.
    }
    \label{tab:prior_section}
\end{table}

\section{Response of $S_8$ and intrinsic alignment to errors in $n(z)$}
\label{sec:response}

We describe the impact of errors in redshift bin estimation on the measurement of cosmological parameters by analyzing multiple mock observations with varying $\delta_z$ or $\sigma_z$, and fixed cosmological and IA parameters.
We use a noiseless data vector for the mock observation to purely focus on the impact of the systematic errors.
Cosmological and intrinsic alignment parameters are fixed to their fiducial values for the mock observation, shown in Table~\ref{tab:prior}.
In the fitted model, we vary the cosmological and intrinsic alignment parameters.
We also vary the redshift shifts $\delta_z$ as nuisance parameters, while the widths are fixed to $\sigma_z=1$ following the standard analysis choice of current Stage-III surveys.
A summary of which parameters are varied in mock observation or analysis is given in Table \ref{tab:response}.
The best-fit is determined as the mode of 100~$\chi^2$ minimizations for each mock observation.

\begin{figure}
	\centering
 		\includegraphics[width=1.00\textwidth]{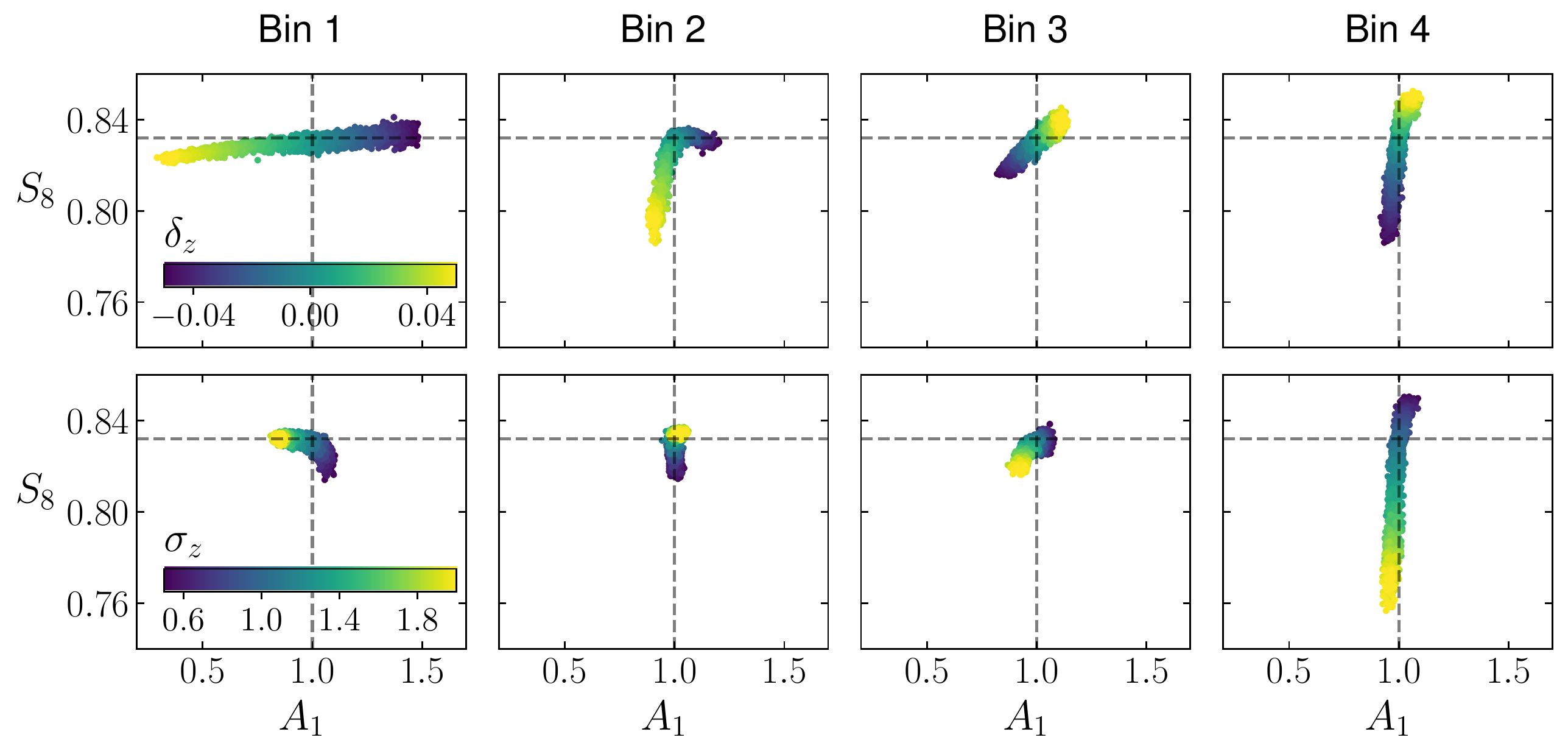}
	\caption{Change in best-fit value of $A_1$ and $S_8$ when varying one redshift parameter at once.
	First row: shift of the whole redshift bin by $\delta_z$.
	Second row: change of width of the whole redshift bin by a factor of $\sigma_z$.
	The different columns correspond to the different redshift bins.
	The parameter that is varied in the most upper left panel is therefore $\delta_{z,1}$.
	The value of $\delta_z$ and $\sigma_z$ in the mock observation are indicated by the color of the dot.
	The dashed lines corresponds to the fiducial values that are used in the analysis.
	We choose $S_8 = 0.832$ (result from Planck \cite{planck_collaboration_planck_2020}) and $A_1 = 1$.
	For this plot, a Stage-III setup using NLA as intrinsic alignment is used.}
	\label{fig:impact_of_1zparam_NLA}
\end{figure}

The impact on the best-fits of $S_8$ and $A_1$ for a Stage-III setup using the NLA model, where $A_1$ is the only free IA parameter, is shown in Figure~\ref{fig:impact_of_1zparam_NLA}.
Errors in low redshift bins are shifting the best-fit on $A_1$, while high redshift bins bias the $S_8$ constraint.
Especially, the $\sigma_z$ errors of the last $z$-bin can heavily bias cosmological constraints, leading to a best-fit of $S_8\approx0.76$ in the most extreme case: for $\sigma_z=2$ we obtain a shift of about $2.5\sigma$ in terms of the expected uncertainty. 

The general trends we observe remain the same when repeating this analysis with a different Stage-III-like setup: We find the same trends when using the KiDS-1000 redshift bins \cite{asgari_kids-1000_2021} and a corresponding covariance matrix.
Using a Stage-IV setup does not change the general trend: a high $\sigma_z$ in the last bins lowers $S_8$, a high $\delta_z$ or $\sigma_z$ in the first bin lowers $A_1$.
We note, however, that the magnitude of this bias is reduced.
This is not due to the fact of the reduced shape noise and cosmic variance since these two have only a negligible impact on the best-fit of the parameter space.
The reduction is due to the higher coverage in redshift which reduces the overlap between the bins and adds additional cosmological information which makes the best-fit more stable to redshift uncertainty.
Nevertheless, comparing the deviation from the true value in terms of the expected standard deviation of such surveys, the impact on Stage-IV surveys would be even larger than for Stage-III.
We find a deviation of close to $6\sigma$ for the most extreme case $\sigma_z=2$ being more than double the deviation of the Stage-III setup.

\begin{figure}
	\centering
		\includegraphics[width=1.00\textwidth]{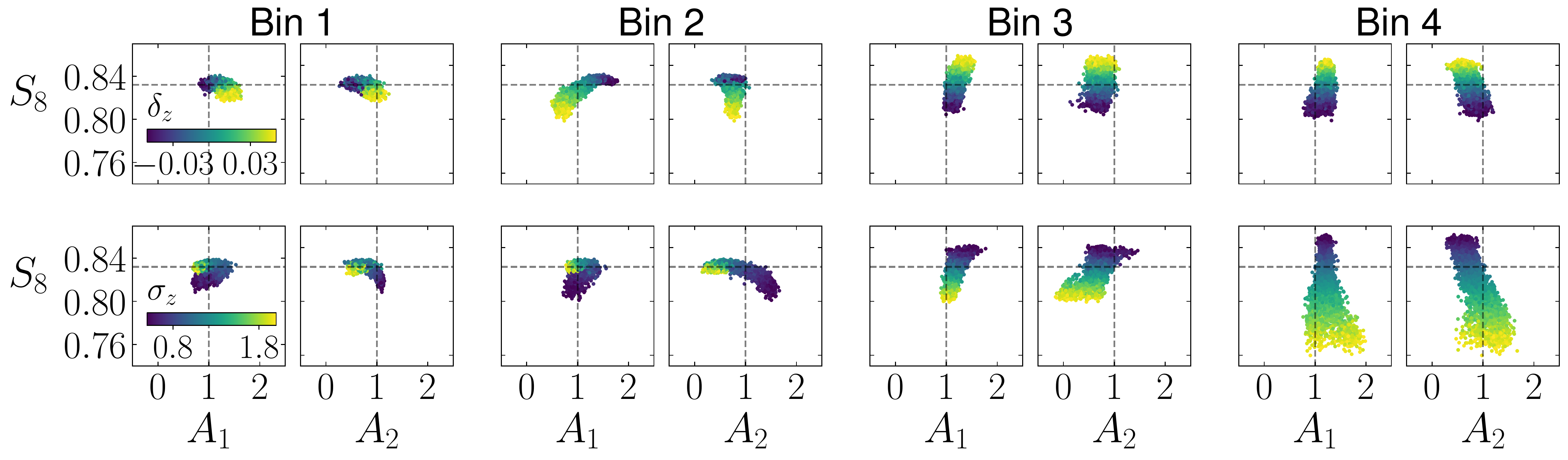}
	\caption{
	Change in best-fit value of $A_1, A_2$ and $S_8$ when varying one redshift parameter at once.
	First row: shift of the whole redshift bin by $\delta_z$. Second row: change of width of the whole redshift bin by a factor of $\sigma_z$. 
	The different columns correspond to the different redshift bin.
	The parameter that is varied in the most upper left panel is therefore $\delta_{z,1}$.
	The value of $\delta_z$ and $\sigma_z$ in the mock observation are indicated by the color of the dot.
	The dashed lines corresponds to the fiducial values that are used in the analysis. 
	We choose $S_8 = 0.832$ (result from Planck \cite{planck_collaboration_planck_2020}) and $A_1 = A_2 = 1$.
	For this plot, a Stage-III setup using TATT as intrinsic alignment model is used.
	Note that further TATT parameters as $\eta_1,\eta_2,b_\mathrm{ta}$ are also varied in the analysis but not shown here.
	We note that some of the plots seem to be slightly misaligned to the true cosmology.
	This is due to projection effects in the multidimensional space.
	Generally, increasing $A_1$ can be partly compensated by lowering $A_2$.
	}
	\label{fig:impact_of_1zparam_TATT}
\end{figure}

We repeat the same analysis with a Stage-III survey using the TATT model to account for intrinsic alignment.
Now $A_1, A_2, \eta_1, \eta_2$ and $b_\mathrm{ta}$ are all varied in the analysis.
The results are shown in Figure \ref{fig:impact_of_1zparam_TATT}.
We find again a strong relation between errors in the redshift bin width $\sigma_z$ of the last bin with the best-fit in $S_8$.
The trends for the lower redshift bins get much more complicated when introducing the additional TATT parameters.

Similarly to the case above, the results do not change much when using a different Stage-III-like setup.
The same is true for the Stage-IV setup: general trends remain the same but the absolute values of maximum biases decrease.
If we weight this decrease by the expected uncertainty of a Stage-IV survey, we again find an increase compared to Stage-III.
We find maximum deviation of around $2.4\sigma$ for Stage-III and about $5.9\sigma$ for Stage-IV.
The reason for the slight decrease in the number of $\sigma$ is due to the increasing uncertainty when using the TATT model due to additional free parameters.

\section{Ignoring $n(z)$ width error: underfitting bias}
\label{sec:ignore}

As shown in Section~\ref{sec:response}, ignoring the error in the redshift bin width can cause biases in~$S_8$ or~$A_1$ due to underfitting.
In this section, we investigate this behavior in more detail by injecting a~$\sigma_z$ error on all redshift bins at the same time, which is not fitted for in the model.
We use a test configuration with the following systematic error in the mock observation
\begin{equation}\label{eq:testsetup}
\begin{aligned}
    &\text{Stage III:}\quad \sigma_{z,1}=0.8,\ \sigma_{z,2}=0.8,\ \sigma_{z,3}=1.2,\ \sigma_{z,4}=1.2,\\
    &\text{Stage IV:}\quad \sigma_{z,1}=1.1,\ \sigma_{z,2}=1.1,\ \sigma_{z,3}=1.1,\ \sigma_{z,4}=1.1,\ \sigma_{z,5}=1.1.
\end{aligned}
\end{equation}
The parameters are chosen in a way that the resulting bias in~$S_8$ is clearly visible, while the~$\sigma_z$ and~$\delta_z$ errors are still in a more or less realistic range.

We use a noiseless mock observation to focus exclusively on the systematics in the redshift estimation and to avoid possible noise-related effects.
We use a standard analysis setup with free cosmological and IA parameters, where the number of free IA parameters depends on the IA model (Table \ref{tab:IAmodels}).
In the fitted model, we use Gaussian priors on $\delta_z$ (Table~\ref{tab:prior}), while $\sigma_z$ is not varied in the analysis and set to $\sigma_z=1$.
The analysis setup of this section is summarized in Table \ref{tab:ign}.

\begin{figure}
         \includegraphics[width=0.495\textwidth,valign=t]{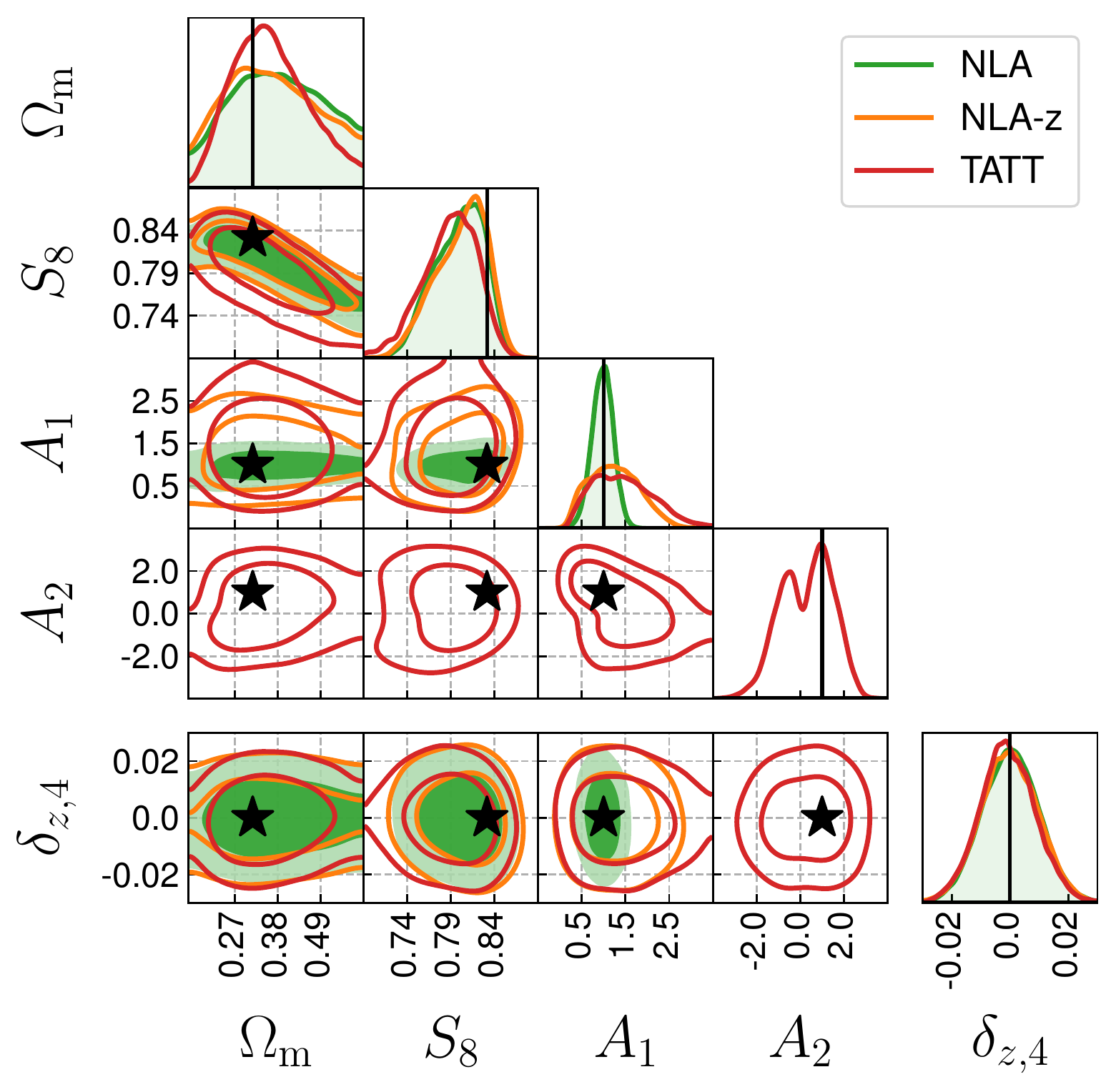}
         \includegraphics[width=0.495\textwidth,valign=t]{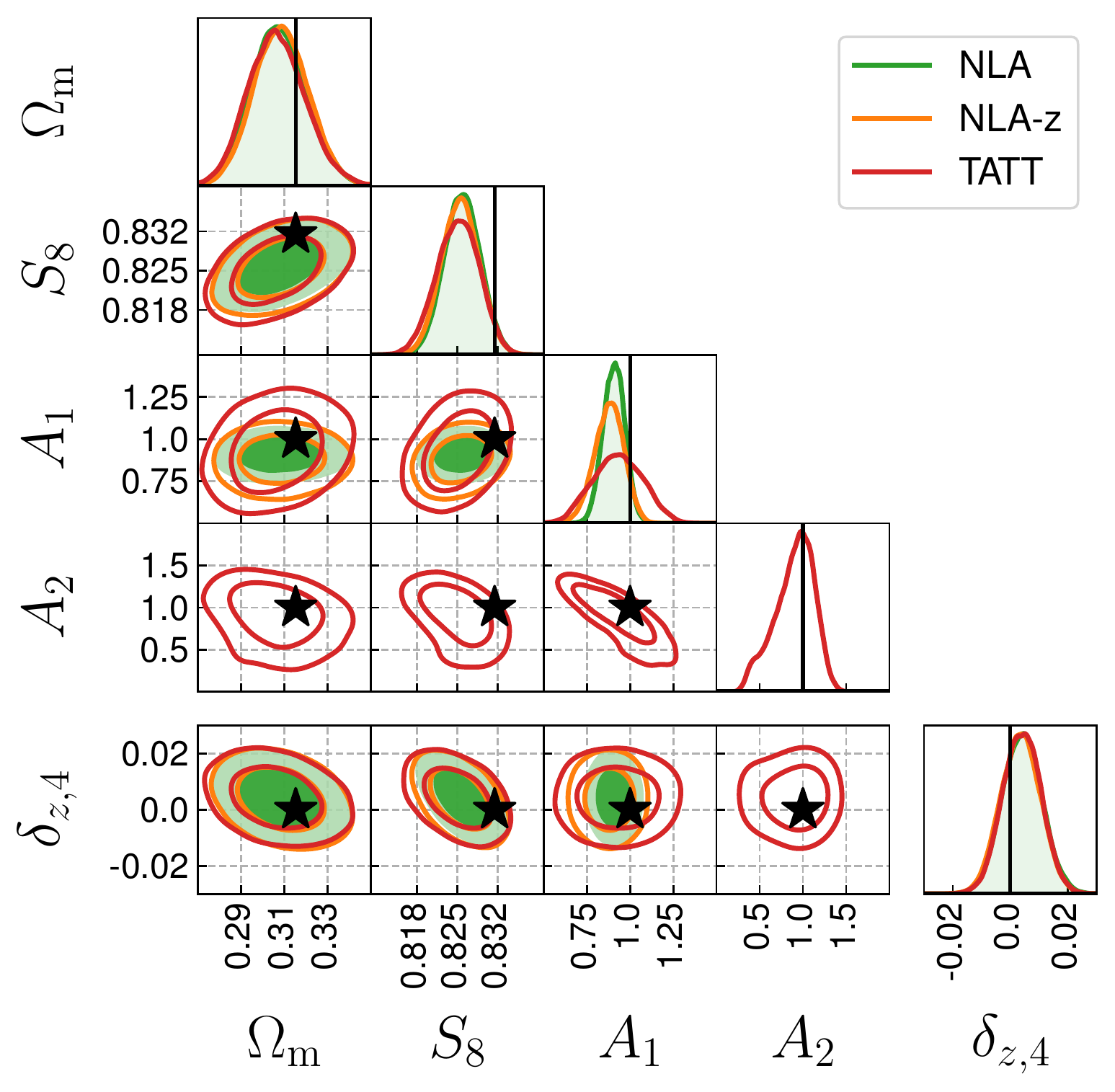}
         \caption{Constraints (68\% and 95\% confidence level) for the Stage-III (left) and Stage-IV (right) configurations for different IA models.
         We inject an error on the redshift bin width given by Equation \ref{eq:testsetup}, but it is ignored in the fitted model, which leads to underfitting. For clarity, we show only a subset of fitted parameters. The full fitted model has 10, 11 or 14 parameters for NLA, NLA-z and TATT, respectively.}
         \label{fig:contours_ignore}
         \vspace{1em}
         \includegraphics[width=0.495\textwidth,valign=t]{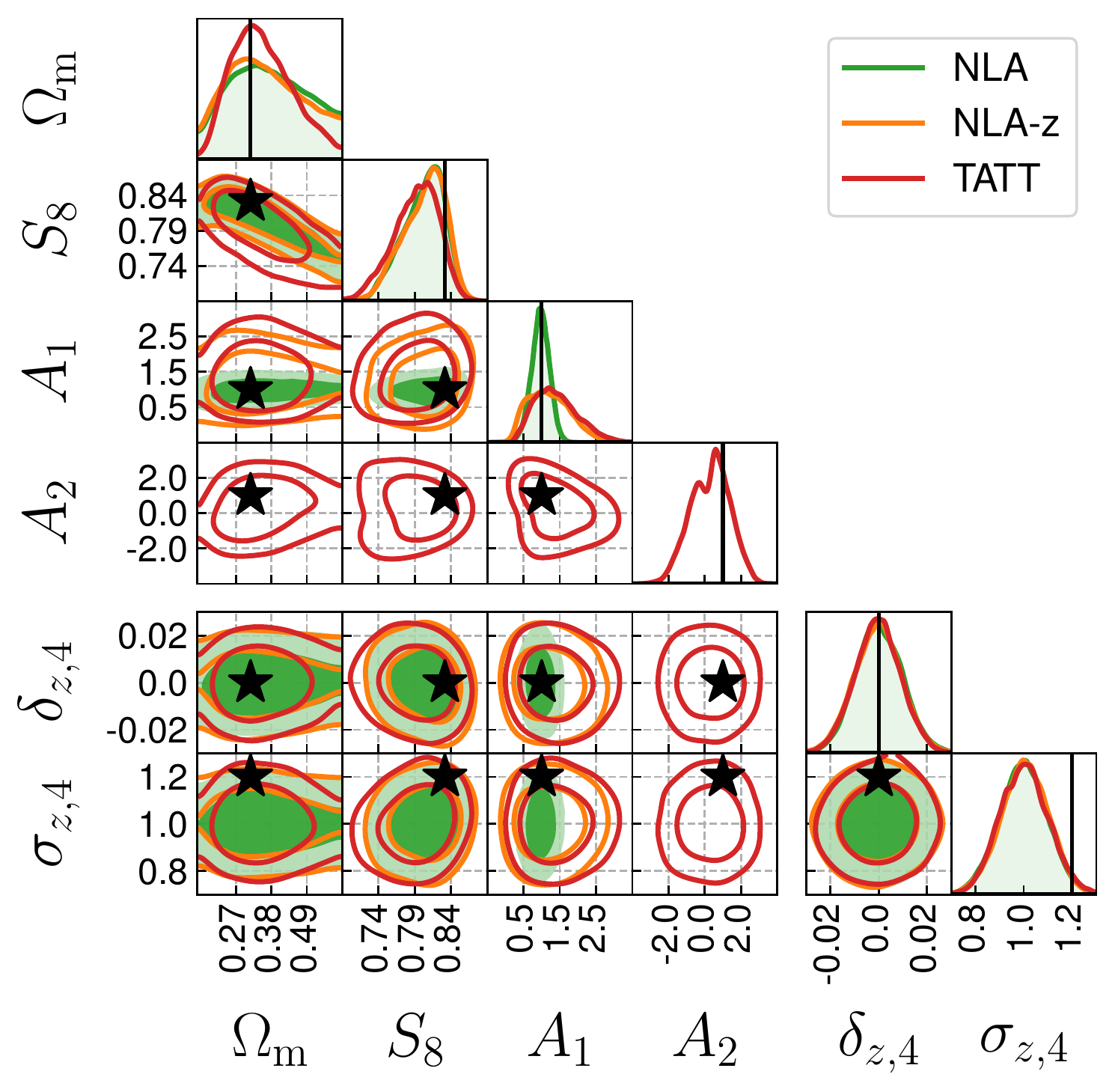}
         \includegraphics[width=0.495\textwidth,valign=t]{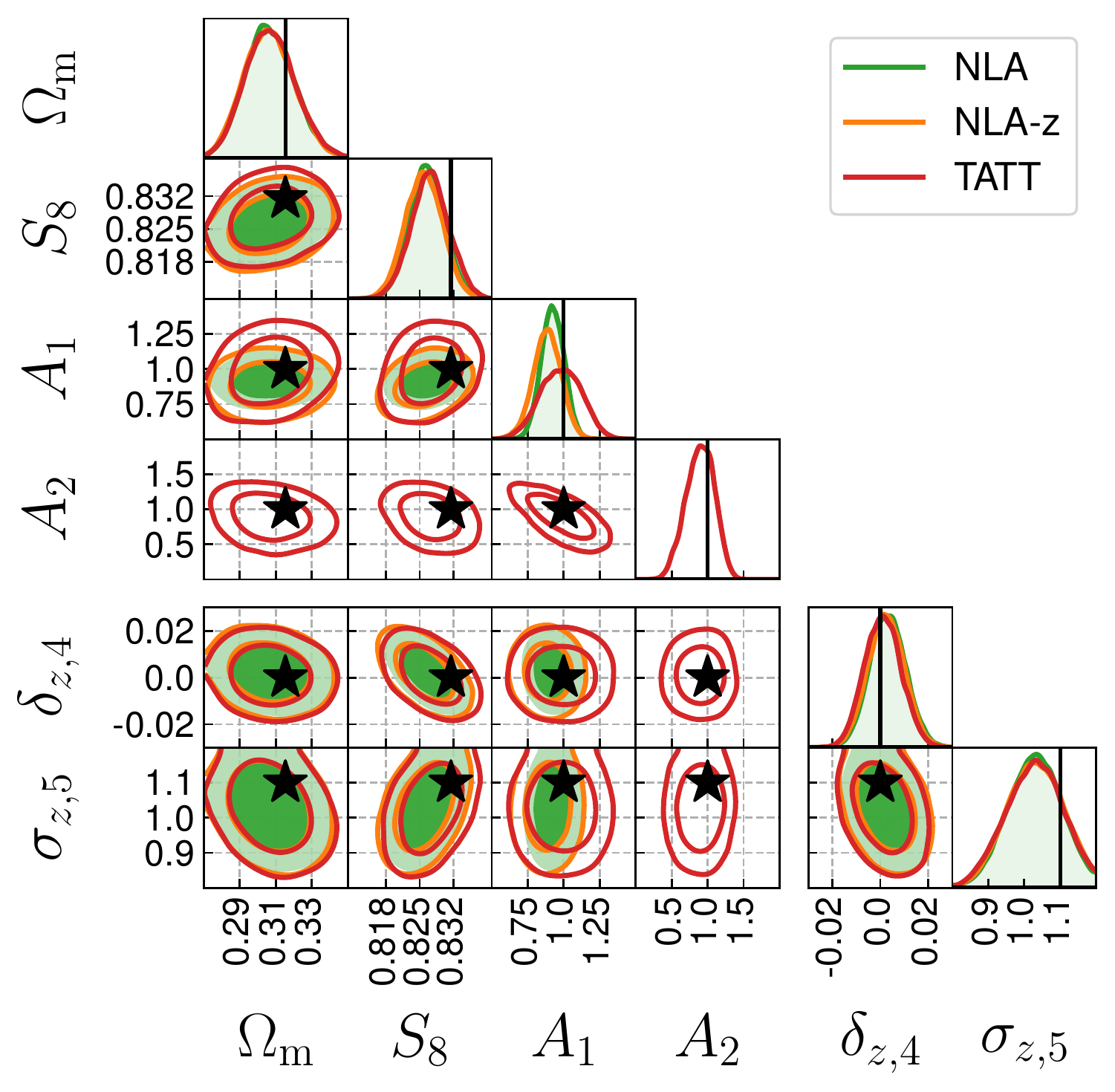}
         \caption{Constraints (68\% and 95\% confidence level) for the Stage-III (left) and Stage-IV (right) configurations for different IA models.
         The redshift bin width error $\delta_z$ is included in the fitted model.
         We inject an error on the redshift bin width given by Equation \ref{eq:testsetup}, which is $2\sigma$ away from the center of the $\sigma_z$ prior for Stage-III and $1\sigma$ away for Stage-IV. For clarity, we show only a subset of fitted parameters. The full fitted model has 11, 12 or 15 parameters for NLA, NLA-z and TATT, respectively. }
         \label{fig:contours_include}
\end{figure}

The resulting posteriors for different IA models are shown in Figure~\ref{fig:contours_ignore}.
For clarity, we plot only the parts of the cosmology and IA parameters, as well as $\delta_{z,4}$, although all other cosmology, IA or redshift parameters were varied in the fitted model.
We see a clear bias in $S_8$ due to the underfitting of the parameter $\sigma_z$ or Stage-III.
Marginalization over $\delta_z$ has no ability to self-calibrate $n(z)$, the Gaussian priors are just perfectly reproduced.
In the Stage-IV setup, the bias in $S_8$ increases relatively to the total uncertainty.
We find slight deviations from the Gaussian priors on $\delta_z$ due to the higher constraining power of such a survey.
This deviation is e.g.\ visible in $\delta_{z,4}$ (see right panel in Figure~\ref{fig:contours_ignore}).
It may be due to the model trying to fit the signal difference caused by the induced $\sigma_z$ error by modifying $\delta_z$.

% TK Removing this sentence, as it it's not clear why we should expect the delta_z to resove this underfitting bias
% Nevertheless, the marginalization over $\delta_z$ does not resolve the underfitting bias.

\section{Including $n(z)$ shape error in the model: sensitivity to priors}
\label{sec:include}
A straightforward step to address the underfitting bias described in Section~\ref{sec:ignore} is to include~$\sigma_z$ in the analysis and marginalize over it with a prior.
In this section we investigate if this approach can alleviate the problem.
We add the $\sigma_z$ parameter to the fitted model and repeat the analysis from Section \ref{sec:ignore} with the same test setup (Equation \ref{eq:testsetup}).
We assume Gaussian priors on~$\sigma_z$ with standard deviation of 0.1.
The analysis is performed with free cosmological, intrinsic alignment and redshift bin error parameters (see Table \ref{tab:prior}).
The analysis setup is summarized in Table \ref{tab:incl}.

The posterior of the Stage-III configuration is shown in the left panel of Figure \ref{fig:contours_include}.
We observe hardly any deviation from the Gaussian prior on the left leading to the more or less same contours as for the case when $\sigma_z$ is ignored.
Repeating the analysis with flat priors on $\sigma_z$, we see that we do not have the power to constrain $\sigma_z$ in a Stage-III survey.

This changes for a Stage-IV survey, where $\sigma_z$ gets slightly constraint and the bias is partly resolved.
The constraints for this case are shown on the right panel in Figure \ref{fig:contours_include}.
For clarity, we show only a subset of fitted parameters.
The full fitted model has 16, 17 or 20 parameters for NLA, NLA-z and TATT, respectively.
We see slight shifts away from the prior for the $\sigma_{z,5}$ parameter, with $\sigma_{z}$ for the other redshift bins look similar.
This leads to a reduction of the bias compared to the case where we ignore $\sigma_z$ in Figure \ref{fig:contours_ignore} (right panel).

Repeating the same setup with flat priors on $\delta_z$ and $\sigma_z$ can again change the results; see Figure \ref{fig:flat_prior} in the appendix for the full chain.
We see that most redshift parameters are at least slightly constrained.
Generally, low redshift bin widths seem to be better constrained by the data.
Including $\sigma_z$ in this noiseless test setup reduces the bias but comes at the cost of significantly increasing the uncertainty by almost a factor of 2 depending on the size of the redshift priors.

Choosing priors for redshift parameters in Stage-IV surveys is therefore a trade-off between reducing biases and increasing the uncertainty.
The same is true for Stage-III, where redshift parameters are very poorly constrained.
Therefore, widening the priors on redshift parameters to reduce the bias leads to an even stronger increase of the uncertainty.
In such a weak information regime with a multi-dimensional parameter space, the errors due to one parameter can be compensated by modifying others, which can lead to biased results; here the error in $\sigma_z$ is compensated by choosing the wrong solution for $S_8$ and other parameters.
Using Gaussian priors that are centered at the wrong value will therefore rather lead to a shift in parameters with flat priors than to self-calibration of the parameter with the Gaussian prior.
In this example, correcting for the wrongly centered Gaussian prior on $\sigma_z$ is easier when shifting cosmological parameter such as $S_8$ then moving away from the Gaussian prior on $\sigma_z$.
The final posterior is therefore sensitive to the priors.
If the information increases and parameters start to be constrained, part of the bias is reduced by deviating from the Gaussian prior.
However, how much the bias is reduced and how much the uncertainty increases still depends on the size of the prior and the deviation from the central prior to the truth.
Stage-IV surveys will therefore still be sensitive to the choice of priors.
\section{Requirements}
\label{sec:req}
As shown in Section~\ref{sec:include}, biases in cosmological parameters due to redshift errors can not be fully avoided by their marginalization due to prior volume effects.
We can quantify the expected uncertainty in the constraints given the size of the prior in the redshift parameters.
The size prior depends on the quality of photometric redshift distribution $n(z)$ calibration.
In this section, we calculate the requirements on $\delta_z$ and $\sigma_z$ for Stage-III and Stage-IV surveys, for two cases: (i) when $\sigma_z$ is \emph{included} and (ii) when it is \emph{ignored} in the analysis configuration.
In this calculation, we assume that the systematic errors due to redshift calibration should not exceed 0.5$\sigma$ of the statistical error of the survey.
We repeat this calculation for different IA models, always assuming that we have the \emph{correct} IA model; there are no errors due to IA model under- or over-fitting.
We calculate requirements on $\sigma_z$ in tandem with requirements on $\delta_z$.
While in the previous sections we find biases for many cosmological and IA parameters, we focus here on $S_8$, since this parameter is of primary interest in weak lensing surveys.
We perform the following computations to determine bias and systematic uncertainty due to redshift errors.

\subsection{Noise-only}
As a first step, we compute the contribution of noise to the bias and uncertainty of $S_8$.
We create 1000~mock observations with noise, leaving the cosmological, IA, and redshift parameters at their fiducial values (i.e.\ no redshift error at all).
This is done for all four IA models (noIA, NLA, NLA-z, TATT).
Each mock observation is then analyzed with flat priors on cosmology and IA, but with fixed values for the redshift bins.
Combining these 1000~MCMC chains and 1000~best-fits, we obtain a stacked chain and a distribution of all best-fits.
We compute the 68\% highest density interval of the stacked chain, which we use as an estimate of the statistical uncertainty of the survey due to noise $\sigma_{\mathrm{n}}[S_8]$. 
The median of all best-fits is used as an estimate of the expected $S_8$.
Comparing this value with the true $S_8$ in the mock observations, we obtain the median bias $b_{\mathrm{n}}[S_8]$ due to noise.

To summarize the results from multiple noise realizations, we use the average posterior described in Equation~\ref{eqn:average_posterior}.
The stacked chains are shown in Figure~\ref{fig:noise} in Appendix~\ref{app:noise}.
The $S_8$ biases for different intrinsic alignment models are given in Table~\ref{tab:noise}.
We expect that noise contributes to the uncertainty budget but does not bias the constraints in a linear model.
This is not true for nonlinear models, and we already find biases due to noise only, especially for the Stage-III setup.
These biases, going up to~$0.5\sigma$, will be investigated in future work.
More details on the noise-only case are given in Appendix~\ref{app:noise}.

\subsection{Noise and redshift systematics}
In a second step, we add noise and redshift systematics to the mock observation and include the redshift parameters in the analysis.
Since we want to vary the uncertainty of the redshift parameters, we have to repeat the setup from the noise-only case several times with different uncertainties of the redshift parameters.
We create a $10\times10$~grid in which we vary $\delta_z$ and $\sigma_z$ so that we get 100~different experimental configurations.
The uncertainty on the shift is varied linearly $\delta_z \in [10^{-5},0.03]$ and the uncertainty on the width parameter $\sigma_z \in [0.01, 0.3]$ for \mbox{Stage-III}.
For Stage-IV, we keep the lower limit the same and halve the upper limits $\delta_z \in [10^{-5},0.015$] and $\sigma_z \in [0.01, 0.15]$.
For each of these 100~setups, we perform an analysis very similar to the noise-only case.
We create 1000~mock observations with cosmological and IA parameters fixed to their fiducial values.
The redshift parameters are drawn from a Gaussian distribution with a width corresponding to the point in the grid; for example, for the setup where the prior width of $\delta_z$ is $0.03$ and the prior width of $\sigma_z$ is $0.3$, in the mock observations $\delta_z$ is varied according to a Gaussian $\mathcal{N}[0,0.03]$, and $\sigma_z$ according to $\mathcal{N}[1,0.3]$.
These mock observations are then analyzed with flat priors for cosmological and IA parameters and Gaussian priors for the redshift parameters.
The width of the Gaussian priors in the analysis correspond exactly to the prior width that is used in the mock observation.
This procedure is performed for four IA models (noIA, NLA, NLA-z, TATT), varying $\sigma_z$ for the \emph{include} case, and once without varying $\sigma_z$ for the \emph{ignore} case.
The~$\sigma_z$ is always varied in the mock observation.

Similar to the noise-only case, we obtain a stacked chain and a distribution of best-fits for each of the 100~setups.
We again compute the highest density interval of the stacked chain and the median of the distribution of best-fits to determine the uncertainty $\sigma_{\mathrm{n+s}}[S_8]$ and the bias $b_{\mathrm{n+s}}[S_8]$ due to noise and redshift systematics.

\begin{table}[t]
    \centering
    \begin{tabular}{lll}
    \toprule
    \textbf{symbol} & \textbf{description} & \textbf{how it is measured}\\
    \midrule
    $b_\mathrm{n}[S_8]$ & $S_8$ bias due to noise & $\hat\mu(\Delta S_8)$ for noise\\
    $b_\mathrm{n+s}[S_8]$ & $S_8$ bias due to noise+signal & $\hat\mu(\Delta S_8)$ for noise+signal\\
    $b_\mathrm{s}[S_8]$ & $S_8$ bias due to redshift systematics & $b_{\mathrm{n+s}}[S_8] - b_{\mathrm{n}}[S_8]$ \\
    \midrule
    $\sigma_\mathrm{n}[S_8]$ & $S_8$ uncertainty due to noise & HDI of stacked noise chains\\
    $\sigma_\mathrm{n+s}[S_8]$ & $S_8$ uncertainty due to noise+signal & HDI of stacked n+s chains\\
    $\sigma_\mathrm{s}[S_8]$ & $S_8$ uncertainty due to redshift systematics & $\sigma_{\mathrm{s}}[S_8] = \sqrt{\sigma_{\mathrm{n+s}}^2[S_8] - \sigma_{\mathrm{n}}^2[S_8]}$ \\
    $\sigma_\mathrm{av}[S_8]$ & average $S_8$ uncertainty & HDI of centered stacked chains\\
    \bottomrule
    \end{tabular}
    \caption{
    Summary of all the bias and uncertainty terms used in this analysis.
    We present for each used bias or uncertainty term a general description of it and how we obtain this parameter.
    We use ''n+s'' for noise and redshift systematics, $\hat\mu$ for the median, $\Delta S_8$ for the difference between best-fit and true $S_8$ and HDI for highest density interval.}
    \label{tab:b_sigma_summary}
\end{table}

\subsection{Deriving the redshift systematics}
In a final step, we compute the uncertainty and bias due to the redshift systematics only.
Using the above results for the noise-only and the noise+systematics case, we can compute the systematic contribution $\sigma_{\mathrm{s}}[S_8]$ as follows. 
\begin{align}
    \sigma_{\mathrm{s}}[S_8] &= \sqrt{\sigma_{\mathrm{n+s}}^2[S_8] - \sigma_{\mathrm{n}}^2[S_8]},  \label{eqn:sigma_and_b} \\
    b_{\mathrm{s}}[S_8] &= b_{\mathrm{n+s}}[S_8] - b_{\mathrm{n}}[S_8],
\end{align}
where we assume that the systematic and the statistical uncertainty are uncorrelated and the uncertainty contributions add in quadrature.
The results for the Stage-III setup are shown in~Figure \ref{fig:req_DES} and for the Stage-IV setup in~Figure \ref{fig:req_LSST}, where the bias $b[S_8]$ is indicated by the color of the dot and the uncertainty $\sigma[S_8]$ by the size of the dot.
The uncertainty is given as the ratio $\sigma_{\mathrm{s}}[S_8]/\sigma_{\mathrm{n}}[S_8]$ between systematic and statistical uncertainty.
A value of $\sigma_{\mathrm{s}}[S_8]/\sigma_{\mathrm{n}}[S_8]>1$ therefore corresponds to a case where the redshift systematics dominate the error budget whereas for $\sigma_{\mathrm{s}}[S_8]/\sigma_{\mathrm{n}}[S_8]<1$ the noise is the main contribution.

The bias is given as $b_{\mathrm{s}}[S_8]/\sigma_{\mathrm{av}}[S_8]$ where the average uncertainty $\sigma_{\mathrm{av}}[S_8]$ is computed as follows.
For each of the 1000~chains per mock observation, we have the corresponding best-fit.
Since we want to know the average uncertainty over the 1000~chains without being affected by systematic biases or noise marginalization, we shift each chain so that the best-fit is centered on the truth.
After centering all chains on the truth, we compute the uncertainty as before to determine $\sigma_{\mathrm{av}}[S_8]$.
Since we center all chains on the same value, we have \mbox{$\sigma_{\mathrm{av}}[S_8]<\sigma_{\mathrm{n+s}}[S_8]$}.
Thus, a value of $b_{\mathrm{s}}[S_8]/\sigma_{\mathrm{av}}[S_8]=0.25$ corresponds to a case where the median $S_8$ measurement differs by $0.25\sigma$ from the true $S_8$ value, in terms of the average uncertainty of an individual analysis.
The different bias and uncertainty terms are summarized in Table \ref{tab:b_sigma_summary}.

\subsection{Requirements for Stage-III surveys}

\begin{figure}

    \begin{subfigure}{\textwidth}
        \centering
        \includegraphics[width=1\textwidth]{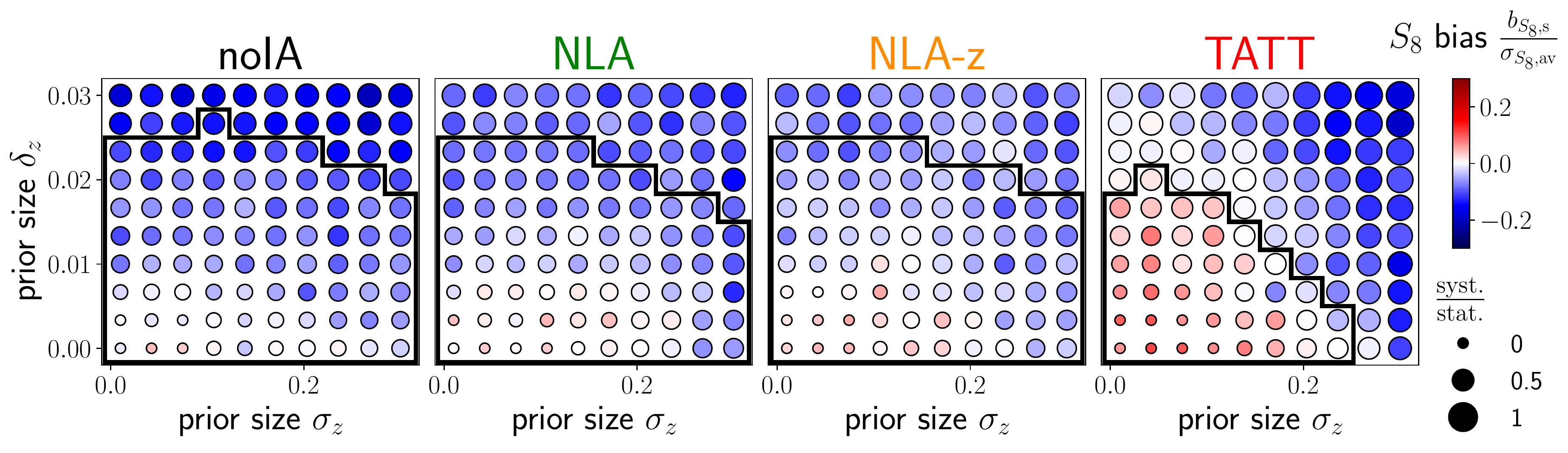}
        \caption{Ignoring $n(z)$ shape error in the model}
        \label{fig:requirements_ignore3} 
    \end{subfigure}
    \begin{subfigure}{\textwidth}
        \centering
        \includegraphics[width=1\textwidth]{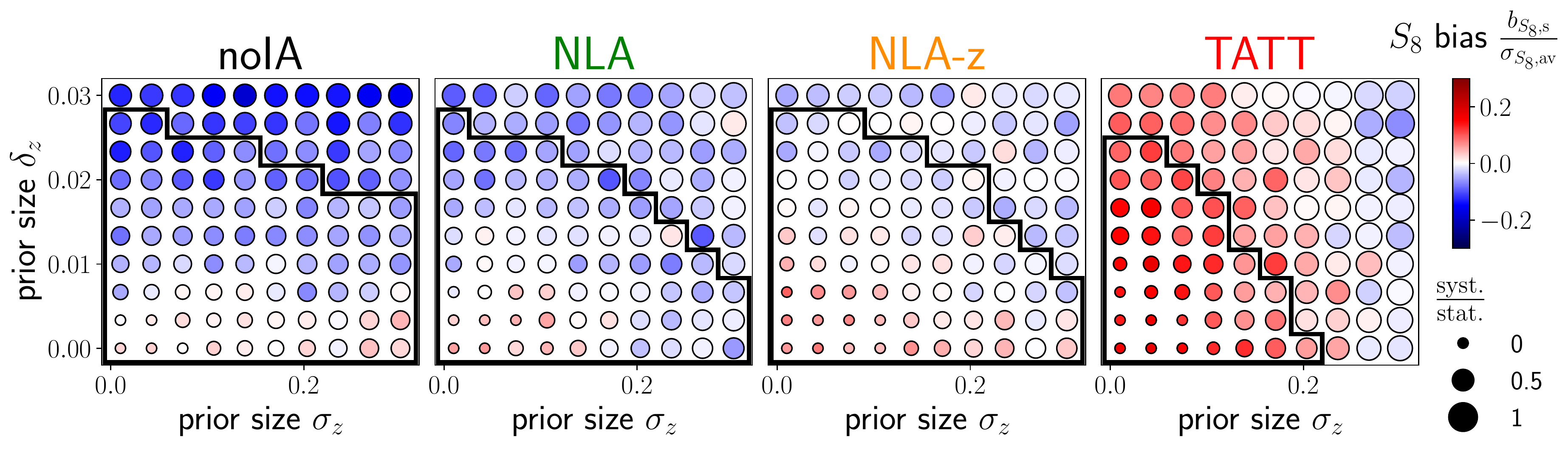}
        \caption{Including $n(z)$ shape error in the model}
        \label{fig:requirements_include3}       
    \end{subfigure}
    \caption{
    Requirements for the Stage-III setup.
    The different dots corresponds to scenarios with varying uncertainty on $\delta_z$ and $\sigma_z$.
    The color of each dot corresponds to the median $S_8$-bias of 1000~mock observations compared the truth.
    The size of each dot corresponds to the ratio between the systematic uncertainty and the statistical uncertainty.
    The scenarios in the black frame fulfill the requirement given by Equation \ref{eq:req}.
    We emphasize that the results presented here are an average over many different noise realisations with varying redshift errors, where the results in Section \ref{sec:ignore} and \ref{sec:include} show one specific setup without noise to provide intuition for the problem.
    }
    \label{fig:req_DES}
\end{figure}

We find rather low biases of $\lesssim0.2\sigma$ for Stage-III.
We generally see that the redshift systematics tend to lead to lower $S_8$.
This is particularly problematic since noise already drives $S_8$ to lower values (see Table \ref{tab:noise}).
This can lead to negative biases of up to $0.6\sigma$ when taking into account the bias of the noise and the redshift systematics.
The dependence on the IA model is also non-trivial:
while in the noise-only case more complex IA models lead to stronger biases (see again Table \ref{tab:noise}), adding redshift systematics seems to have the opposite effect.
More complex IA models seem more likely to push $S_8$ back up, while less complex IA models lead to even lower $S_8$ values.
Comparing the setup where we include $\sigma_z$ in the analysis with the setup where we ignore $\sigma_z$, we find only small differences for noIA, NLA and NLA-z.
This is not unexpected, since we have seen in Section \ref{sec:include} that marginalization over $\sigma_z$ does not have a large effect for the Stage-III setup.
For TATT, we find that including $\sigma_z$ reduces the (negative) bias for large uncertainties but increases the (positive) bias for lower uncertainties in $\delta_z$ and $\sigma_z$.
Although the bias due to redshift systematics can be positive, the survey -- in combination with the bias caused by noise -- is still biased toward lower $S_8$ for all combinations of redshift uncertainties.

Increasing the redshift errors lead to higher systematic uncertainties, as expected.
This increase is stronger for more complex IA models which can be easily explained by the increase in parameters that are varied in the analysis.
We also find a slight increase when including $\sigma_z$ in the analysis.
This is due to the higher number of free parameters in the n+s analysis which increases the uncertainty.
This effect is stronger than a possible decrease in uncertainty due to more correct modelling.
Since Stage-III surveys have a very high noise component, the redshift systematics dominate the error budget only for very high redshift uncertainty and complex IA models.

We report requirements on the redshift calibration for $\delta_z$ and $\sigma_z$.
The requirements are determined by the statistical power of the survey in two ways:
(i) the results should not be biased by more than a small fraction of the calculated uncertainty,
(ii) the systematic uncertainty should not dominate the total calculated uncertainty.
For this paper we set the following requirement thresholds:
\begin{equation}
    \label{eq:req}
    \begin{aligned}
    b_{\mathrm{s}}[S_8] &< 0.25 \cdot \sigma_{\mathrm{av}}[S_8], \\
    \sigma_{\mathrm{s}}[S_8] &< 0.5 \cdot  \sigma_{\mathrm{n}}[S_8].
    \end{aligned}
\end{equation}
where $ \sigma_{\mathrm{av}}[S_8]$ is the average total uncertainty of a survey at a given stage, including the contributions of the noise and systematic marginalization,  $\sigma_{\mathrm{s}}[S_8]$ is the systematic uncertainty (Equation~\ref{eqn:sigma_and_b}),  and $\sigma_{\mathrm{n}}[S_8]$ is the uncertainty due to the statistical noise only.
We emphasize that these are rather conservative thresholds since we only consider redshift systematics.
If $0.5\cdot \sigma_{\mathrm{n}}[S_8]$ were to be found, the scope for other systematics, such as shear calibration, would be greatly limited.

Configurations that fulfill the requirements for Stage-III are shown with a black frame in Figure~\ref{fig:req_DES}.
The driving source of the requirement is the systematic uncertainty; the bias requirement is fulfilled for each dot.
Generally, current weak lensing surveys are clearly in this frame with some additional margin.
We also find that, with the current data, the inclusion of~$\sigma_z$ does not significantly increase the tolerance for redshift errors.

\subsection{Requirements for Stage-IV surveys}
\begin{figure}
    \begin{subfigure}{\textwidth}
        \centering
        \includegraphics[width=1\textwidth]{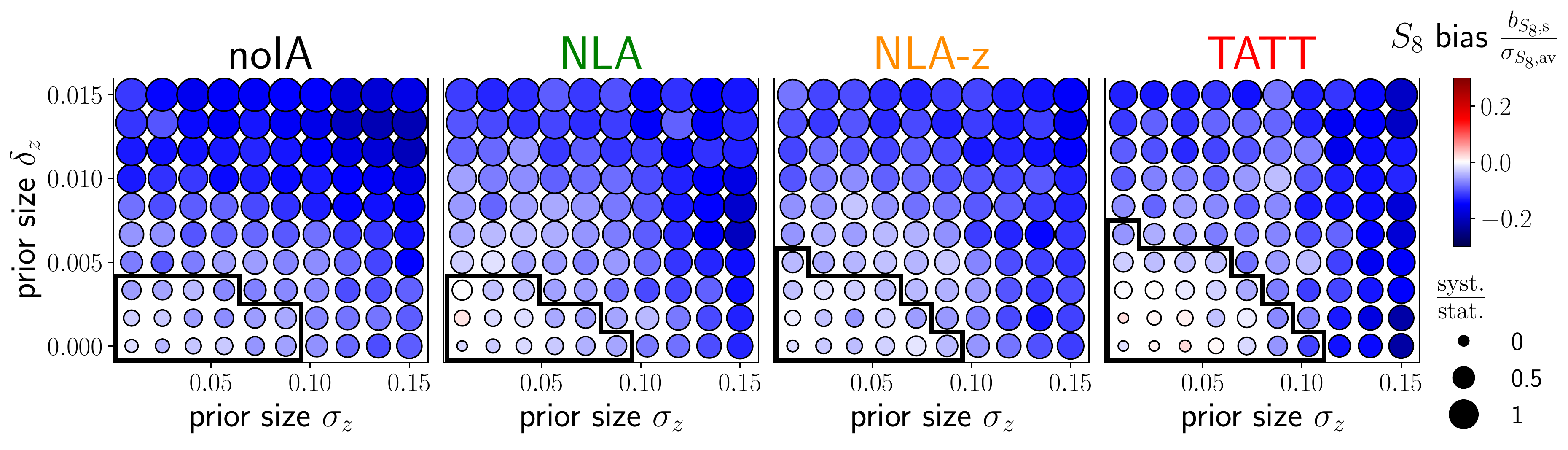}
        \caption{Ignoring $n(z)$ shape error in the model}
        \label{fig:requirements_ignore4} 
    \end{subfigure}
    \begin{subfigure}{\textwidth}
        \centering
        \includegraphics[width=1\textwidth]{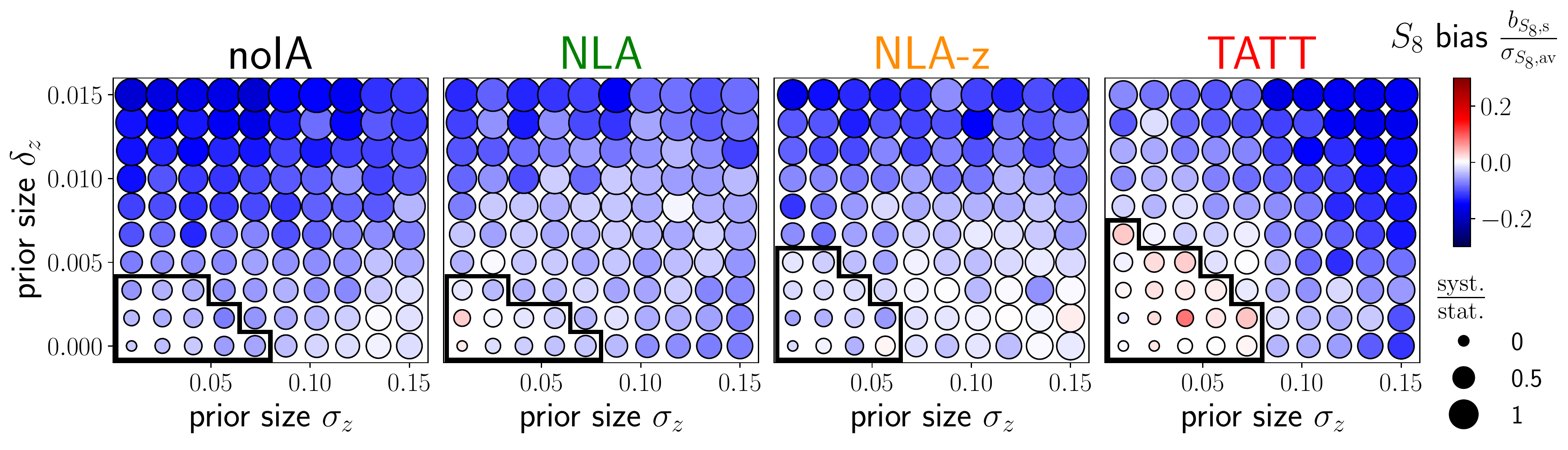}
        \caption{Including $n(z)$ shape error in the model}
        \label{fig:requirements_include4}       
    \end{subfigure}
    \caption{
    Requirements for the Stage-IV setup.
    The different dots corresponds to scenarios with varying uncertainty on $\delta_z$ and $\sigma_z$.
    The color of each dot corresponds to the median $S_8$-bias of 1000~mock observations compared the truth.
    The size of each dot corresponds to the ratio between the systematic uncertainty and the statistical uncertainty.
    The scenarios in the black frame fulfill the requirement given by Equation \ref{eq:req}.
    Note that the axis limits correspond to half of the limits of Figure \ref{fig:req_DES}.
    We emphasize that the results presented here are an average over many different noise realisations with varying redshift errors, where the results in Section \ref{sec:ignore} and \ref{sec:include} show one specific setup without noise to provide intuition for the problem.
    }
    \label{fig:req_LSST}
\end{figure}

Biases, as compared to the average uncertainty, increase when considering the Stage-IV survey.
Redshift systematics lead to lower $S_8$ for all combinations of prior widths if $\sigma_z$ is ignored in the analysis.
Since the noise contribution alone (see Table \ref{tab:noise}) already pushes $S_8$ slightly down, this leads to even lower $S_8$ values.
If $\sigma_z$ is included in the analysis, we find combinations of prior widths tending to higher values, especially for less complex IA models and large prior widths.
We find maximum biases of up to $0.7\sigma$ due to redshift systematics if using the same grid as in Stage-III, which is much larger than the bias from the noise only.
This reduces to a maximum bias of about $0.2\sigma$ in the zoomed grid shown in Figure \ref{fig:req_LSST}, but it is still clearly larger than the noise bias.
Therefore, the analysis is biased towards lower $S_8$ when combining the bias from noise and redshift systematics and $\sigma_z$ is ignored.
When $\sigma_z$ is included in the model, the survey can also be biased towards higher $S_8$ for less complex IA models and large prior width.

The uncertainties increase smoothly with the prior width and the complexity of the IA model.
Compared to Stage-III, we no longer find large differences in the level of systematic uncertainty between the IA models.
Nevertheless, the systematic uncertainty of the survey still increases for more complex IA models; note that we always consider the correct IA model and do not examine biases due to incorrect IA.
But the same is true for the statistical uncertainty, which includes the contribution of additional IA parameters.
As we consider the ratio between systematic and statistical uncertainty, larger systematic uncertainty is compensated by larger statistical uncertainty if the increase of the two uncertainties is similar.
This is the case for the Stage-IV setup.

The decrease in absolute values of the statistical uncertainty leads to a more dominant role of the redshift systematics compared to Stage-III.
We find $\sigma_{\mathrm{s}}[S_8]/\sigma_{\mathrm{av}}[S_8]\approx 3.2$ for the most extreme cases when using the same grid as in Stage-III.
For the zoomed grid, we find $\sigma_{\mathrm{s}}[S_8]/\sigma_{\mathrm{av}}[S_8]\approx 1.9$ for the most extreme case.
Redshift systematics dominate the error budget in large parts of the grid.

The scenarios for Stage-IV, in which the requirements are fulfilled, are shown by the black frame in Figure \ref{fig:req_LSST}.
For our choice of requirements, including $\sigma_z$ neither improves bias nor uncertainty.
This is due to the fact that we require a very high precision in both redshift parameters in order to fulfill our requirements.
In this regime, including $\sigma_z$ mainly leads to complications in the MCMC and best-fit determination and does not significantly improve the model; therefore, ignoring $\sigma_z$ leads to very similar or even slightly better results.
If we would tolerate much higher uncertainties, including $\sigma_z$ leads to better results especially in terms of bias due to the better modelling.

Generally, including $\sigma_z$ will not change the cosmological constraints significantly as seen in the similarity of the two setups in Figure \ref{fig:req_LSST}.
However, this does not imply that we can tolerate arbitrarily high uncertainty on $\sigma_z$.
To fulfill the requirement, we need the uncertainty on $\sigma_z$ to be $\lesssim0.1$ for all IA models.

\subsection{Application to past and planned surveys}
\label{sec:req_currents}

If the pipeline presented above is applied to past and planned surveys, the exact numbers of the requirements will change depending on the survey specifications.
However, estimating a realistic uncertainty on $\delta_z$ and $\sigma_z$ of current Stage-III and upcoming Stage-IV surveys, we get an idea of how realistic it is to fulfill these requirements and we get a rough idea of the expected systematic uncertainty.

Current Stage-III surveys typically only consider uncertainties on $\delta_z$ in their analyses.
KiDS-1000 reports their estimate for $\delta_z$ from 0.0087 to 0.0118 with rather larger uncertainty for the low redshift bins \cite{asgari_kids-1000_2021}.
HSC reports photo-z uncertainty between 0.0135 and 0.0383 with rather larger uncertainty for higher redshift bins \cite{hikage_cosmology_2019}.
DES-Y3 reports the spread in mean redshift between 0.015 and 0.018 with no significant redshift dependence \cite{amon_dark_nodate}.

They do not report any estimate of their uncertainty on the $n(z)$ width.
An approach to avoid specifiying the $\sigma_z$ uncertainty is the \textsc{hyperrank} method \cite{cordero_dark_2021}.
They account for the redshift uncertainty by marginalizing over different discrete redshift realisations.
Since it uses different realisations and not only shifted versions of the same redshift distributions, it accounts for all types of redshift distribution errors.
Consistent to our results, they find no notable deviations when using the more complex model (in this case \textsc{hyperrank}) for the redshift uncertainties when analyzing the DES-Y3 data \cite{cordero_dark_2021}.
However, this may change for upcoming surveys with higher precision.

A rough estimate of the width errors can be inferred when comparing the redshift bins from \cite{wright_kidsviking-450_2019}, where DIR was used for redshift calibration, and \cite{wright_kidsviking-450_2020}, where SOM was used.
We find that using different methods to measure the photometric redshifts can lead to discrepancies in $\sigma_z$ of up 0.2.
However, since the two redshift samples only partially overlap, a direct comparison is difficult.

Assuming a $\sigma_z$ uncertainty of 0.1 and $\delta_z$ uncertainty of 0.01, we can directly compute the expected uncertainties and biases for different IA models.
They are given in Table \ref{tab:bias_and_uncertainty3}.
Already here, the systematic uncertainties take up to a third of the statistical uncertainty.
However, the requirement we set in this work is fulfilled with some margin.

\begin{table}[t]
    \centering
        \begin{tabular}{l|llll}
            \toprule 
            \multicolumn{5}{c}{\textbf{ignoring $\sigma_z$ in the model}} \\
      \cmidrule(lr){1-5}
            & \multicolumn{1}{c}{noIA} & \multicolumn{1}{c}{NLA} & \multicolumn{1}{c}{NLA-z} & \multicolumn{1}{c}{TATT}\\
            $b_{\mathrm{s}}[S_8]/\sigma_{\mathrm{av}}[S_8]$ & -0.051 & -0.027 & 0.019 & 0.039\\
            $\sigma_{\mathrm{s}}[S_8]/\sigma_{\mathrm{n}}[S_8]$ & 0.26 & 0.27 & 0.25 & 0.33\\
            
            \midrule
            \multicolumn{5}{c}{\textbf{including $\sigma_z$ in the model}} \\
      \cmidrule(lr){1-5}
            & \multicolumn{1}{c}{noIA} & \multicolumn{1}{c}{NLA} & \multicolumn{1}{c}{NLA-z} & \multicolumn{1}{c}{TATT}\\
            $b_{\mathrm{s}}[S_8]/\sigma_{\mathrm{av}}[S_8]$ & -0.067 & -0.007 & 0.004 & 0.125\\
            $\sigma_{\mathrm{s}}[S_8]/\sigma_{\mathrm{n}}[S_8]$ & 0.31 & 0.29 & 0.26 & 0.36\\
            \bottomrule
        \end{tabular}
    \caption{Expected median bias and level of systematic uncertainty for redshift errors corresponding to current Stage-III surveys.}
    \label{tab:bias_and_uncertainty3}
\end{table}

Following the science requirements document of LSST \cite{the_lsst_dark_energy_science_collaboration_lsst_2018}, we can estimate the uncertainty of the redshift distribution.
The current requirements on $\delta_z$ are given by $0.001(1 + z)$.
No specific requirement is given for $\sigma_z$.
For the highest redshift bins going up to $z > 2$, we expect to be close to the thresholds of our requirement, depending on the uncertainty of $\sigma_z$.
However, we would like to emphasize that this is not necessarily true for all Stage-IV setups.
This analysis should be repeated with the exact setup of the survey to get a survey-specific forecast, since the results can vary for different covariance matrices or $\ell$-ranges (see Appendix~\ref{app:req_dep}).
We also consider only the same prior width for all redshift bins.
If future surveys have very different uncertainties on these parameters, for example high precision for low redshift and rather large uncertainties for high redshift bins, one should account for this when repeating the type of analysis presented in this work.
But the fact that we are already close to the thresholds with the expected uncertainty for $\delta_z$ shows that a systematic uncertainty due to redshift errors of about half the statistical uncertainty is a realistic scenario.

\section{Can the interplay between IA modelling and redshift shape error cause the $S_8$-tension?}
\label{sec:lowS8}
Errors in the redshift width estimation can lead to biases on cosmological constraints.
In this section, we investigate if an interplay between IA modelling and errors in the redshift estimation can explain the $S_8$-tension between Planck and weak lensing measurements.
For that, we analyze multiple mock observations without noise and determine the best-fit of $S_8$ for each of them.
We vary cosmological and intrinsic alignment parameters, depending on the IA model, as well as $\delta_z$, the redshift bin width parameter $\sigma_z$ is fixed.
See Table \ref{tab:prior} for details.

The mock observation is computed as follows.
The cosmological parameters are fixed to their fiducial values and intrinsic alignment parameters are varied.
We also vary the two redshift parameters $\delta_z$ and $\sigma_z$; we use flat priors with a range of $\delta_z \in [-0.02,0.02]$ and $\sigma_z \in [0.5,2]$, differently to the priors used in previous sections and listed in Table~\ref{tab:prior}.
Since we are concerned about the $S_8$-tension arising in the most recent Stage-III surveys, we use the Stage-III setup in this section.
We perform the analysis both for NLA and TATT as intrinsic alignment model.
The summary of the parameters and how they are varied in mock observation and analysis is given in Table~\ref{tab:lowS8}.

We pick all the scenarios with a best-fit of $S_8<0.78$ to find possible combinations of parameters that lead to a low $S_8$.
We find many such combinations, some of them exclusively due to the width in the last bin (see also Section \ref{sec:response}) or due to underfitting the intrinsic alignment contribution (see Appendix \ref{app:underfitting_IA}).
In Figure \ref{fig:S8_tension}, we show scenarios that illustrate how intrinsic alignment and redshift estimation errors are coupled.
We vary in both cases only redshift bin widths $\sigma_{z,i}$ and IA parameters that are also included in the model, i.e.\ $A_1$ for NLA and $A_1, A_2, \eta_1, \eta_2, b_\mathrm{ta}$ for TATT.
Again, no IA underfitting was considered in this study.
All combinations of parameters that lead to $S_8<0.78$ are then plotted as a 2-D histogram.
Note that this is not a probability distribution -- it is a set of scenarios that lead to low $S_8$.
The color in Figure \ref{fig:S8_tension} corresponds to the relative density of these low $S_8$ scenarios compared to all scenarios.
For NLA, we see that in the most extreme case of $\sigma_{z,4}\sim1.5$ and $A_1\sim 4$, we always find such a low $S_8$.
Therefore the low $S_8$ density is 1.
For TATT, the highest density can be found for very high $\sigma_{z,4}$ and high $A_1$ and corresponds to roughly 10\% of all scenarios in this region of the parameter space.

For NLA, the most low-$S_8$ scenarios are created for errors in the width of the highest redshift bin.
Nevertheless, there is also a clear trend for the first two redshift bins, where compressed bins lead to a low $S_8$.
But all these low $S_8$ scenarios only exists if the intrinsic alignment amplitude $A_1>0$.
The higher $A_1$ is, the less extreme values are needed in $\sigma_z$ to get a low $S_8$.
We emphasize that $A_1$ is included in our analysis and we would naively expect that the value of $A_1$ in the mock observation should not matter.
But the errors in the redshift estimation are propagated through the intrinsic alignment modelling and depending on the amplitude, the final bias in $S_8$ can be increased or decreased by the intrinsic alignment.
High IA amplitude makes the impact of redshift width errors larger: this effect is due to the fact that redshift width errors have a larger impact on the IA power spectrum than on the cosmic shear power spectrum, as illustrated in Figure \ref{fig:concept}.

For TATT, the trends for the redshift bin widths are very similar as in the NLA case.
The IA parameters are also coupled with redshift errors and can amplify or reduce the bias.
Since the TATT model contains five free parameters, the trends are not as clear as in the NLA model.
We notice, however, that there are more scenarios with complicated IA/$z$ dependencies that can lead to low $S_8$.

For both scenarios, adding non-zero errors in mean redshift lead to more possibilities for a low $S_8$.
The same is true when we add the TATT parameters to the NLA analysis and therefore underfit the IA of the mock observation.
Each of these points could fully explain the $S_8$ tension if the assumed errors in the mock observation are actually made by current surveys.
It has to be noted that we also find many scenarios that would lead to a high $S_8$ or to the true $S_8$ even though many errors are assumed in the mock observation.
The solutions presented here belong to a tail of a distribution of best-fits.
If we use tighter priors on the different parameters in the mock observation, which we consider more realistic for an actual survey, the low $S_8$ scenarios account for only about 7.4\% of all scenarios in the case where we fit with NLA (including underfitting IA) and 1.3\% of all scenarios in the case where we fit with TATT.
These tighter priors are: $A_1 \in [-1,2], \ A_2 \in [-1,2], \ \eta_i \in [-2,2], \ b_\mathrm{ta} \in [0.5,1.5], \ \delta_{z,i} \in [-0.01,0.01], \ \sigma_{z,i} \in [0.8,1.2]$.

We therefore conclude that errors in redshift estimation and intrinsic alignment are unlikely the sole explanation of the $S_8$ tension, but not an impossible one.
Nevertheless, these results show clearly how different errors can lead to unexpected couplings and that especially errors on the width of the redshift distribution could lead to large biases in cosmological parameters.

Repeating this analysis with the Stage-IV setup shows similar trends.
Especially the width of the last redshift bin has a large impact on the $S_8$ constraint.
We also discover couplings between IA and redshift errors similar to the Stage-III case.
But the range in which $S_8$ varies decreases drastically due to the higher constraining power.
If a Stage-IV survey will measure $S_8\approx0.76$, we could exclude that the systematic errors described here could be the sole reason for the $S_8$ tension.

\begin{figure}
	\centering
         \centering
         \includegraphics[width=0.38\textwidth,valign=c]{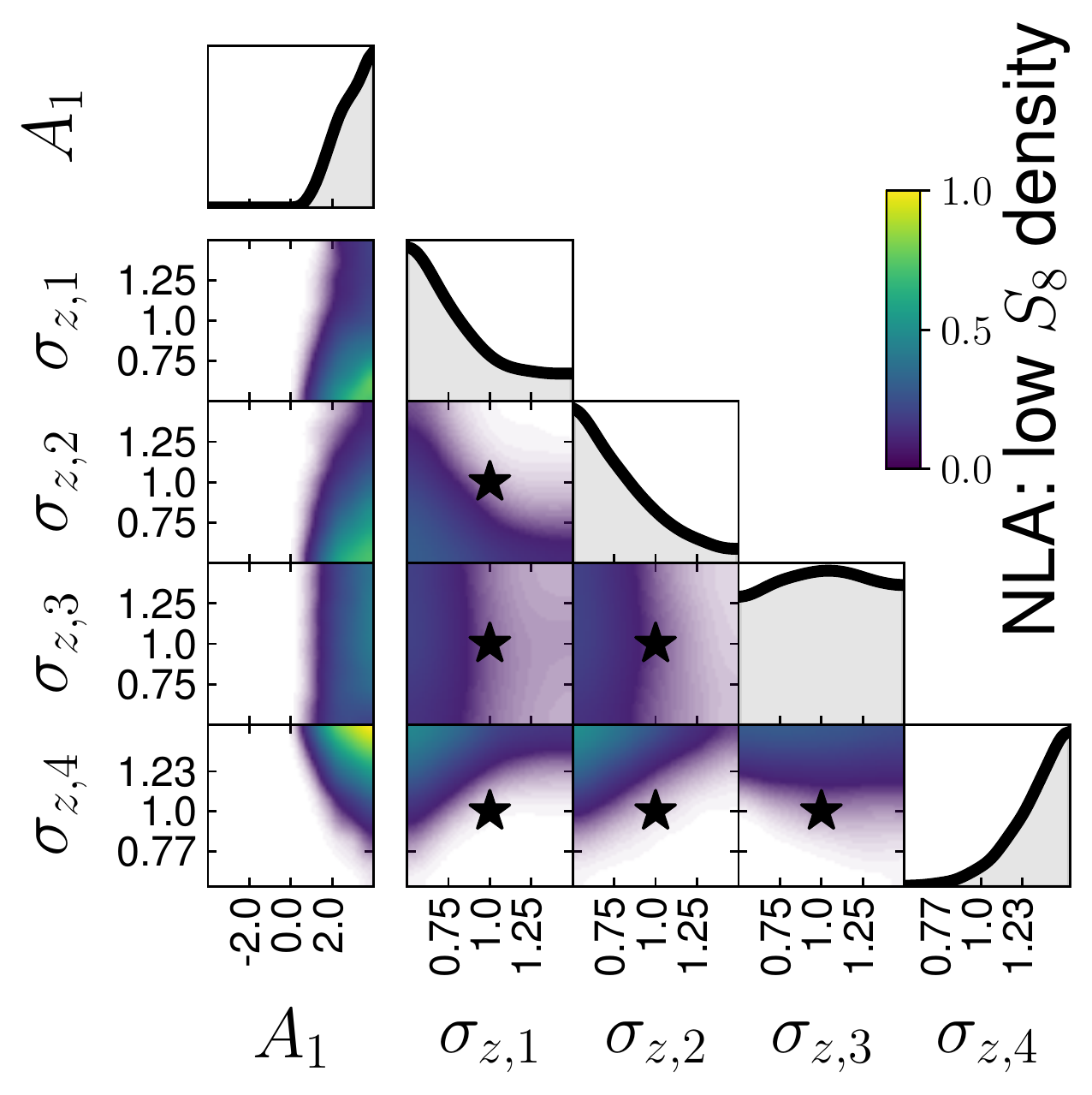}
         \centering
         \includegraphics[width=0.61\textwidth,valign=c]{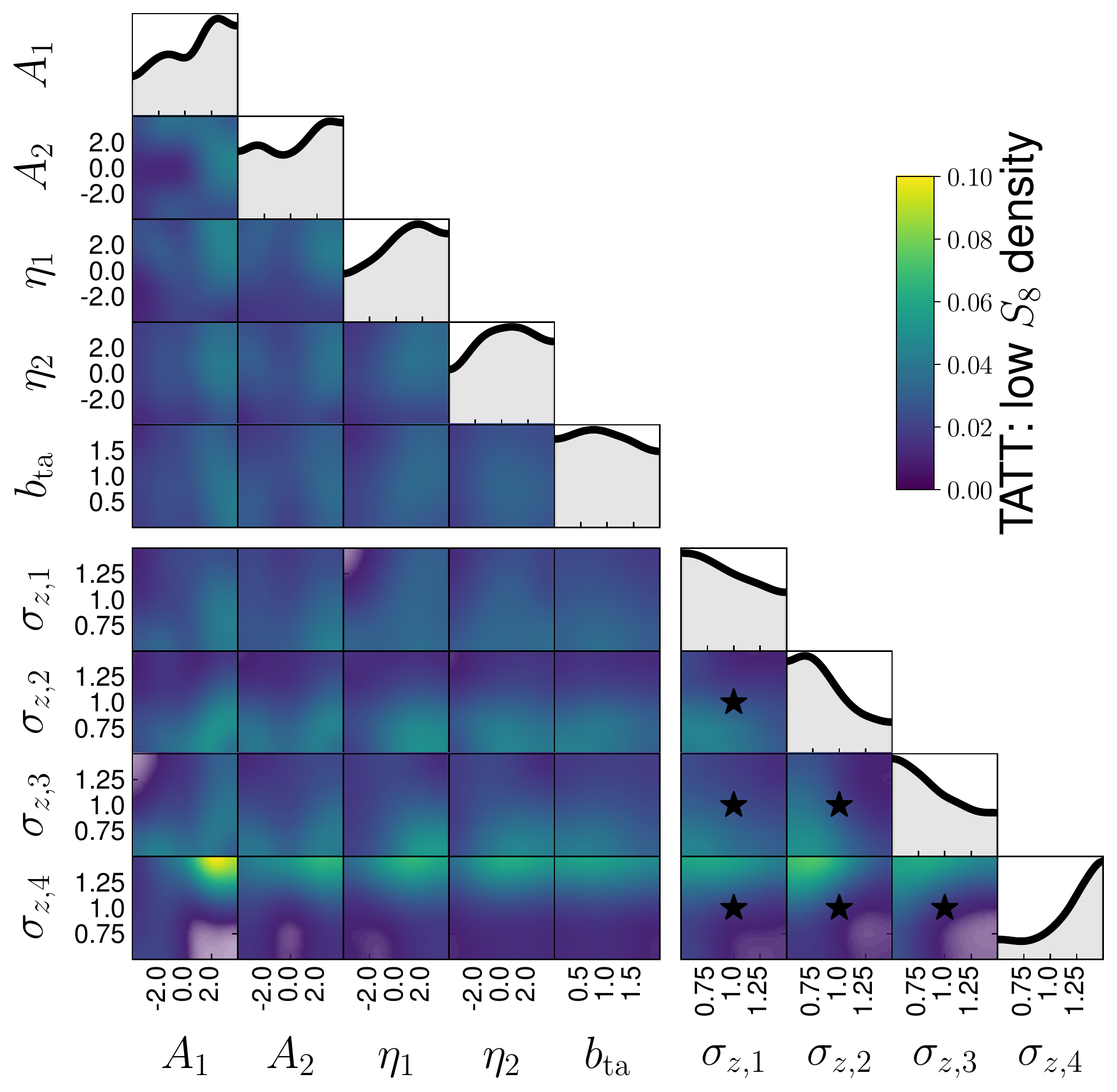}
	\caption{
	Density plots of combinations of parameters that lead a low $S_8$ value in a Stage-III survey.
	We fix cosmological parameters of the mock observation to the values from Planck \cite{planck_collaboration_planck_2020} and vary intrinsic alignment and redshift bin width.
	Each mock observation is then analyzed and a best-fit of $S_8$ is determined.
	All set of parameters that lead to low $S_8<0.78$ are saved.
	The distribution of these sets is shown in these density plots.
	The colorbar corresponds to the relative density of low $S_8$ scenarios compared to all scenarios.
	}
	\label{fig:S8_tension}
\end{figure}

\section{Conclusions}
\label{sec:conclusions}
In this work, we investigate how redshift estimation errors and intrinsic alignment (IA) modelling affect cosmological constraints of weak lensing surveys.
We use the parameters $\delta_z$ and $\sigma_z$ to describe redshift errors, where $\delta_z$ describes shift of the entire redshift bin and $\sigma_z$ changes the width of a redshift bin.
We use three IA models with increasing complexity: NLA, NLA-z and TATT.
We find that both $\delta_z$ and $\sigma_z$ can bias cosmological results if only $\delta_z$ is considered in the analysis, as in most recent weak lensing analyses.
When we combine such redshift errors with different IA models and different fiducial values of the IA parameters, we see that these parameters are coupled.
For example, certain combinations of IA parameters make it much more likely to measure a value of $S_8$ that is too low compared to the truth than others.
For NLA, for example, this is the case the higher the IA amplitude $A_1$ is.

A naive solution to avoid biases due to underfitting of the width errors is to include the width parameter $\sigma_z$ in the analysis.
However, since the constraining power on the redshift parameters is quite low, marginalizing $\sigma_z$ does not necessarily help to reduce such biases.
The results become very dependent on the choice or priors, as the final value is driven by the prior volume effects.

We compute requirements on $\delta_z$ and $\sigma_z$ for current and future surveys for different IA models.
We assume perfect knowledge of the IA model and the size of the redshift uncertainty.
Note that the latter does not mean that the values of each redshift parameter is known but only the size of the Gaussian deviation from the fiducial value.
We find no improvement when including $\sigma_z$ in the analysis for a Stage-III setup.
This is due to the fact that Stage-III data is not powerful enough to constrain the redshift parameters.
Statistical noise accounts for most of the error budget for both bias and uncertainty.
However, assuming an uncertainty for $\delta_z$ of 0.01 and for  $\sigma_z$ of 0.1, we find that redshift systematics account for between a quarter and a third of the uncertainty just due to noise.
The bias, on the other hand, is relatively small, especially when comparing to the bias due to noise of up to $0.5\sigma$.

We determine requirements on $\delta_z$ and $\sigma_z$ simultaneously.
For this purpopse, we require that the systematic bias to be smaller than a fourth of the average total uncertainty: \mbox{$b_{\mathrm{s}}[S_8] < 0.25\sigma_{\mathrm{av}}[S_8]$}, and that the systematic uncertainty is less than half the statistical uncertainty: $\sigma_{\mathrm{s}}[S_8] < 0.5\sigma_{\mathrm{n}}[S_8]$.
Assuming a current precision of $\delta_z = 0.00\pm0.01$ and $\sigma_z = 1.0 \pm 0.1$, we find this requirement to be fulfilled, with some additional margins.

Stage-IV data will greatly increase the precision of cosmic shear measurements and are expected to achieve a greatly improved calibration of redshifts.
Therefore, including $\sigma_z$ in the analysis will make a difference in the posterior distribution of cosmological parameters.
Due to the decreasing impact of noise, the systematic part of the uncertainty grows as well as biases.
Including $\sigma_z$ tends to reduce the fractional bias while keeping the uncertainty approximately constant.
These biases can be as high as $0.7\sigma$ in the most extreme case.
It is also possible that systematic uncertainty due to redshift errors could dominate compared the noise contribution.
To fulfill the requirements on $\delta_z$ and $\sigma_z$, we need a much higher precision in the redshift estimation than for Stage-III.
However, current estimates of the uncertainty for Stage-IV just fulfill the requirements.
We therefore expect that redshift systematics alone could take up to half of the statistical uncertainty for Stage-IV surveys.

We investigate whether this effect could cause the $S_8$-tension between weak lensing and CMB measurements.
Although we find some realistic scenarios than can explain this tension, they involve rather complicated interdependencies between parameters.
We conclude that the IA and $n(z)$ bias interplay is unlikely to be the sole explanation of the tension.

In this work, we used generic Stage-III and Stage-IV survey redshift bins; we recommend to repeat this analysis with exact specifications of current of upcoming surveys, as the results depend on the shape of the $n(z)$.
To simplify such an analysis, we publish the repository \texttt{refrigerator}\footnote{\url{https://cosmo-gitlab.phys.ethz.ch/cosmo_public/refrigerator}} (REdshiFt RequIrements GEneRATOR) with all the pipelines that were used in this work.

A powerful way of breaking parts of the degeneracy between lensing and IA is the probe combination of lensing and clustering 3x2pt (for recent results using this approach, see \cite{des_collaboration_dark_2022, heymans_kids-1000_2021}).
Extending our pipeline from cosmic shear to a 3x2pt setup would shed light on the impact of redshift distribution errors in such a setup.
Since the IA signal plays a less prominent role in such an analysis, we expect a weaker dependence between different IA models and systematic uncertainty and bias.

We further note that, in the current setup, we assume a best-case scenario where we have perfect knowledge of redshift uncertainty and IA model.
Neither is the case in practice, which may further bias cosmological constraints.
We show how IA underfitting could bias the results in the Appendix.
We also note that we only consider redshift and IA systematics while disregarding others.
Including other systematics will further increase the systematic uncertainty, so it is clear that the treatment of systematics will be a major challenge for upcoming surveys.
\acknowledgments

We thank Janis Fluri, Simran Gurdasani, Beatrice Moser, Alexander Reeves, Raphael Sgier and Dominik Z\"urcher for helpful discussions and Uwe Schmitt for informatics support.
We acknowledge the support of Euler Cluster by High Performance Computing Group from ETHZ Scientific IT Services that we used for most of our computations.

We used functionalities provided by \texttt{numpy} \cite{van_der_walt_numpy_2011}, \texttt{scipy} \cite{virtanen_scipy_2020}, \texttt{arviz} \cite{kumar_arviz_2019}, \texttt{sklearn} \cite{pedregosa_scikit-learn_2018} and \texttt{matplotlib} \cite{hunter_matplotlib_2007} for this work.
Job arrays were submitted using \texttt{esub-epipe} \cite{zurcher_cosmological_2021,zurcher_dark_2022}.
Corner plots were created with \texttt{trianglechain}.\footnote{\url{https://cosmo-gitlab.phys.ethz.ch/cosmo_public/trianglechain/}}
We used \texttt{emcee} \cite{foreman-mackey_emcee_2013} to run our MCMC chains.

\bibliography{references}
\appendix
\section{Setup configurations}
\label{app:setup}
Here, we describe the two setups that are used in this work in more detail.

\subsection{Redshift distribution}
The two redshift distributions are shown in Figure \ref{fig:zsigma} and are created as follows.
We fit each redshift bin of the DES-Y3 analysis \cite{amon_dark_nodate} with a Smail-type distribution given by
\begin{equation}\label{eq:smail}
    n(z) = z^\alpha \exp\left(-\left(\frac{z}{z_0}\right)^\beta\right).
\end{equation}
Although a Smail-type distribution is normally used to describe the total redshift distribution, it provides a very accurate fit to the DES-Y3 bins outperforming other tested fitting functions.
The numerical values of the parameters are given in Table \ref{tab:smail}.

The bins of the Stage-IV survey are created for LSST-Y10 using the numbers given in \cite{the_lsst_dark_energy_science_collaboration_lsst_2018}.
The total redshift distribution is given by a Smail-type distribution with $\alpha = 2, \beta = 0.68$ and $z_0=0.11$.
The resulting distribution is cut into five equally populated bins.
We assume a redshift dependent photometric scatter of $0.05(1+z)$.
\begin{table}[t!]
    \centering
    \begin{tabular}{lccc}
    \toprule
        \textbf{Bin} & $\alpha$ & $\beta$ & $z_0$ \\
        \midrule
         1 & 1.99 & 1.44 & 0.20 \\
         2 & 3.46 & 2.34 & 0.39 \\
         3 & 6.03 & 3.60 & 0.66 \\
         4 & 3.53 & 4.49 & 1.03 \\ 
         \bottomrule 
    \end{tabular}
    \caption{Values for Smail-parameters (see Equation~\ref{eq:smail}) for the redshift bins of the Stage-III survey.}
    \label{tab:smail}
\end{table}

\subsection{$\ell$ binning}
We use different $\ell$ binning schemes for Stage-III and Stage-IV.
We use a $\ell$-range of [100,1000] for Stage-III and bin linearly using 20 bins.
For Stage-IV, we expand the $\ell$-range to [20,3000] and use square-root spaced binning with 50~bins.

\subsection{Covariance matrix}
We need an estimation of the covariance matrix in order to get an accurate mock inference.
This estimation is done using simulations (as described in Section \ref{sec:covmat}).
Survey-specific parameters of these simulations are the survey area, the shape noise component $\sigma_\varepsilon$ and the galaxy number density $n_\mathrm{eff}$ per bin.
The numbers are inspired by DES-Y3 \cite{amon_dark_nodate} for Stage-III and LSST \cite{the_lsst_dark_energy_science_collaboration_lsst_2018} for Stage-IV and are given in Table \ref{tab:covmat}.

\begin{table}[t]
    \centering
    \begin{tabular}{llll}
    \toprule
    \textbf{Survey} & \textbf{Area} & \textbf{shape noise component $\sigma_\varepsilon$} & \textbf{galaxy number density $n_\mathrm{eff}$} \\
    \midrule
    Stage-III & 4143 & $[0.243, 0.262, 0.259, 0.301]$ & $[1.476, 1.479, 1.484, 1.461]$ \\
    Stage-IV & 14300 & $[0.26,0.26,0.26,0.26,0.26]$ & $[5.4,5.4,5.4,5.4,5.4]$ \\
    \bottomrule
    \end{tabular}
    \caption{Parameters used to compute the covariance matrix for the different survey setups.}
    \label{tab:covmat}
\end{table}
\section{Fast $C_\ell$ emulator \chaoshammer}
\label{app:chaoshammer}

Fast prediction of the angular power spectrum given a set of input parameters (cosmology, intrinsic alignments, redshift errors), is the key to exploring the problem presented in the paper. 
To achieve this, we create a fast emulator, dubbed \chaoshammer, based on shallow neural networks. 
\chaoshammer is trained on the full, high precision $C_\ell$ training set of $[\theta,C_\ell]$ pairs generated \pycosmo, where $\theta$ is the input parameter set, the dimensionality of which depends on the model (see Table~\ref{tab:dimensions}).

\begin{table}[ht]
\begin{subtable}[t]{1\textwidth}
    \centering
    \begin{tabular}{c|cccccc}
    \toprule
    & \textbf{NLA} & \textbf{NLA $\sigma_z$} & \textbf{NLA-z} & \textbf{NLA-z $\sigma_z$} & \textbf{TATT} & \textbf{TATT $\sigma_z$} \\ \hline
    \textbf{Stage-III} & 10 & 14 & 11 & 15 & 14 & 18 \\
    \textbf{Stage-IV} & 11 & 16 & 12 & 17 & 15 & 20 \\
    \bottomrule
    \end{tabular}
    \caption{Dimensionality}
    \label{tab:dimensions}
\end{subtable}

\bigskip

\begin{subtable}[t]{1\textwidth}
    \centering
    \begin{tabular}{c|cccccc}
    \toprule
    & \textbf{NLA} & \textbf{NLA $\sigma_z$} & \textbf{NLA-z} & \textbf{NLA-z $\sigma_z$} & \textbf{TATT} & \textbf{TATT $\sigma_z$} \\ \hline
    \textbf{Stage-III} & 2 & 2 & 2 & 2 & 50 & 100 \\
    \textbf{Stage-IV} & 2&2&2&2&2&2 \\
    \bottomrule
    \end{tabular}
    \caption{Number of samples in millions for training}
    \label{tab:nsamples}
\end{subtable}

\bigskip

\begin{subtable}[t]{1\textwidth}
    \centering
    \begin{tabular}{c|cccccc}
    \toprule
    & \textbf{NLA} & \textbf{NLA $\sigma_z$} & \textbf{NLA-z} & \textbf{NLA-z $\sigma_z$} & \textbf{TATT} & \textbf{TATT $\sigma_z$} \\ \hline
    \textbf{Stage-III} & 10 & 10 & 10 & 10 & 14 & 15 \\
    \textbf{Stage-IV} & 10 & 10 & 10 & 10 & 10 & 10 \\
    \bottomrule
    \end{tabular}
    \caption{Batch size in $\log_2$}
    \label{tab:batch_size}
\end{subtable}

\bigskip

\begin{subtable}[t]{1\textwidth}
    \centering
    \begin{tabular}{c|cccccc}
    \toprule
    & \textbf{NLA} & \textbf{NLA $\sigma_z$} & \textbf{NLA-z} & \textbf{NLA-z $\sigma_z$} & \textbf{TATT} & \textbf{TATT $\sigma_z$} \\ \hline
    \textbf{Stage-III} & $99.83\%$ & $99.68\%$ & $99.60\%$ & $99.23\%$ & $99.26\%$ & $98.56\%$ \\
    \textbf{Stage-IV} & $99.95\%$ & $99.91\%$ & $99.90\%$ & $99.81\%$ & $99.62\%$ & $99.33\%$ \\
    \bottomrule
    \end{tabular}
    \caption{Reached accuracy}
    \label{tab:accuracy}
\end{subtable}
\caption{Specifications of the different emulators used in this work.}
\label{tab:emulator}
\end{table}

When using no intrinsic alignment (model ``noIA''), the NLA emulator is used with $A_1=0$. For more details of the models, see Table~\ref{tab:prior}. This appendix describes the construction of the \chaoshammer emulator, its training, and its evaluation.

We begin with creating a grid of training points in the $\theta$ prior parameter space using the Halton low discrepancy sequence.
This set contains an order of 2~to 100~million training points, depending on the model.
Then, we run \pycosmo for each of these points, using 1~core per 1000~points on the Euler cluster. 
\pycosmo generates the tomographic angular power spectra $C_\ell$ (auto and cross), which are then binned into 20~bins in range $\ell \in [100,1000]$ for Stage III and 50~bins in range $\ell \in [20,3000]$ for Stage IV.
Then, the concatenated vector of auto and cross $C_\ell$ is stored.

To learn the function of $C_\ell(\theta)$ over the entire prior range, we use a shallow neural network.
The network takes $\theta$ as input and outputs the tomographic $C_\ell$.
We use 3~hidden layers, each with 512~units.
We build it with the \texttt{Tensorflow} \cite{tensorflow_developers_tensorflow_2021} package and use the stochastic gradient descent optimizer \textsc{Adam} \cite{kingma_adam_2017}.
The loss function is a simple squared L2 norm: $\mathcal{L}=||C_\ell-C_\ell^{\rm{emu}}||_2^2$.
We use a variable batch size indicated in Table \ref{tab:batch_size} and train it for 200 epochs.

In order for the results of our experiments to be accurate, the emulator must satisfy strict precision requirements on its $C_\ell$ prediction.
We define the accuracy in terms of median absolute deviation on the fractional difference between the \pycosmo and \chaoshammer full $C_\ell$ vector predictions. We measure that across the entire prior space, using the training set.
We aim for sub-precision accuracy, which should be sufficient for current and upcoming lensing surveys \cite{schneider_matter_2016, martinelli_euclid_2021}.

We train eight emulators. Details on dimensionality, number of samples, batch size and reached accuracy are given in Table~\ref{tab:emulator}.
Models with larger number of parameters and wider priors need more training points to maintain the accuracy across the entire high-dimensional parameter space.
After the emulators are trained, they generate the $C_\ell$ prediction very fast, in $\approx$ \si{1}{ms}.

We do not consider over-fitting in our models and, for the main training runs, we do not split the sample into training and reserved test sets.
We found that not to be necessary as we have millions of training points, covering the parameter space very densely.
However, we did check if the error level on previously unseen new $\theta$ parameters is of the same order as on the ones in the training set.
We found that to be satisfied. 
\section{Response to underfitting of IA}
\label{app:underfitting_IA}
\begin{figure}
	\centering
        \includegraphics[width=\textwidth]{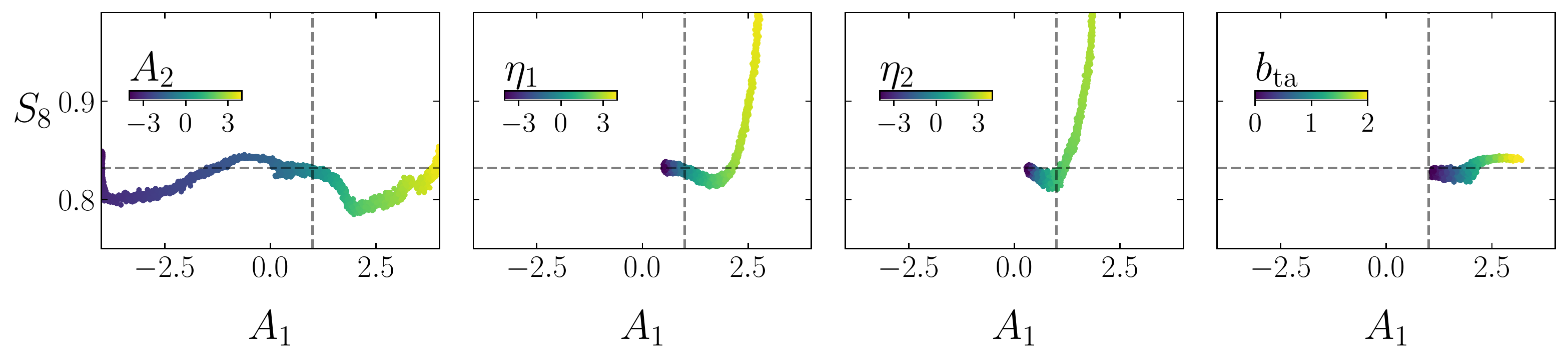}
	\caption{Impact of underfitting different IA parameters. Different TATT parameters are varied in the mock observation while keeping the others fixed and then analyzed using the NLA model. Since we set $A_2=0$ for the fiducial parameter, varying $\eta_2$ would have no impact at all. To still see an effect, we set $A_1=0, A_2=1$ for the $\eta_2$ scenario.}
	\label{fig:impact_of_1IAparam}
\end{figure}

Here, we present the response of cosmological constraints to underfitting of the intrinsic alignment.
We use the NLA model for the analysis but the TATT model for the mock observation.
We present the results for a Stage-III setup although the impact on a Stage-IV setup are very similar.
The parameters are fixed at $A_1=1$ and $A_2=\eta_i=b_\mathrm{ta}=0$ for the fiducial setup.
Then, we vary each parameter separately and compute the best fit for each mock observation.
This is shown in Figure \ref{fig:impact_of_1IAparam}.
Since $\eta_2$ has no impact on the cosmic shear power spectrum if $A_2=0$, varying $\eta_2$ would have no effect on the best-fit of $A_1$ and $S_8$.
To at least get an idea of the general evolution, we set $A_2=1$ for the $\eta_2$ scenario.
To avoid a too large values for $A_1$ in the analysis, we set $A_1=0$.

The impact of $A_2$ is mainly shifting the constraints on $A_1$ -- a higher $A_2$ in the mock observation also leads to a higher $A_1$ in the analysis.
Nevertheless, there is also some effect on $S_8$, varying in the range from about 0.85 to 0.8.
The two redshift dependence parameters $\eta_i$ show both very similar and very extreme trends.
There is some small bias on $A_1$ and $S_8$ for negative and low $\eta_i$, for very high $\eta_i$ the constraints for $S_8$ blow up extremely and hit our prior limit of $S_8=1$ for the very high values close to $\eta_i=4$.
The source bias parameter $b_\mathrm{ta}$ has mainly impact $A_1$ pushing it to higher values.

\begin{figure}
	\centering
        \includegraphics[width=\textwidth]{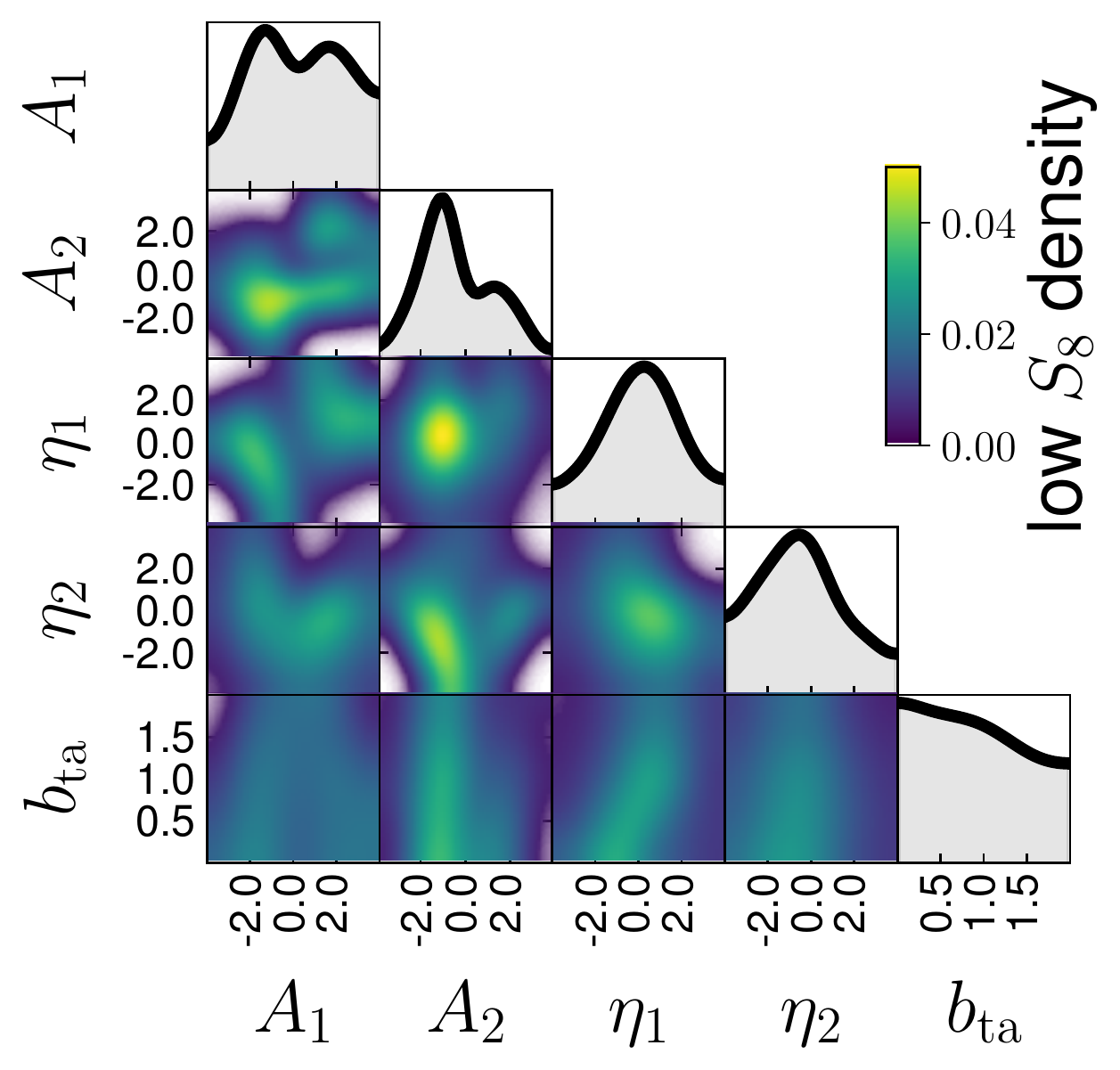}
	\caption{
	Density plots of combinations of parameters that lead to a low $S_8$ in a Stage-III survey assuming a NLA model while the mock observation assumes TATT.
	Mock observations were done with TATT with variable IA parameters and $S_8=0.832$.
	The analysis is done using the NLA model.
	The scenarios presented here correspond to the scenarios where we find a best-fit of $S_8<0.78$.
	}
	\label{fig:impact_of_IAparams}
\end{figure}

If only varying one parameter at the time, only $\eta_i\neq0$ can heavily bias the constraints on $S_8$.
If we vary all IA parameters together with flat priors with the range given in Table \ref{tab:prior}, different combinations can play together and lead to quite strong biases in $S_8$.
Figure \ref{fig:impact_of_IAparams} shows a density plot comparable to the ones shown in Section \ref{sec:lowS8} for this case.
We analyse a TATT Universe with the NLA model, and the combinations shown in Figure \ref{fig:impact_of_IAparams} lead to a low $S_8<0.78$.
As clearly visible there are certain combinations of amplitudes and corresponding redshift dependence that would favor such a low $S_8$ in the final analysis.
Nevertheless, looking at the total distribution of best-fits of $S_8$ over the whole prior range, these scenarios are rather in the tail corresponding to the lowest $11\%$.
Generally, the distribution is rather skewed towards higher $S_8$ which is not surprising looking at Figure \ref{fig:impact_of_1IAparam} where high values of $\eta_i$ lead to a very high $S_8$.
\section{Flat priors on redshift parameters}
\label{app:flat_priors}
\begin{figure}
    \centering
    \includegraphics[width=\textwidth]{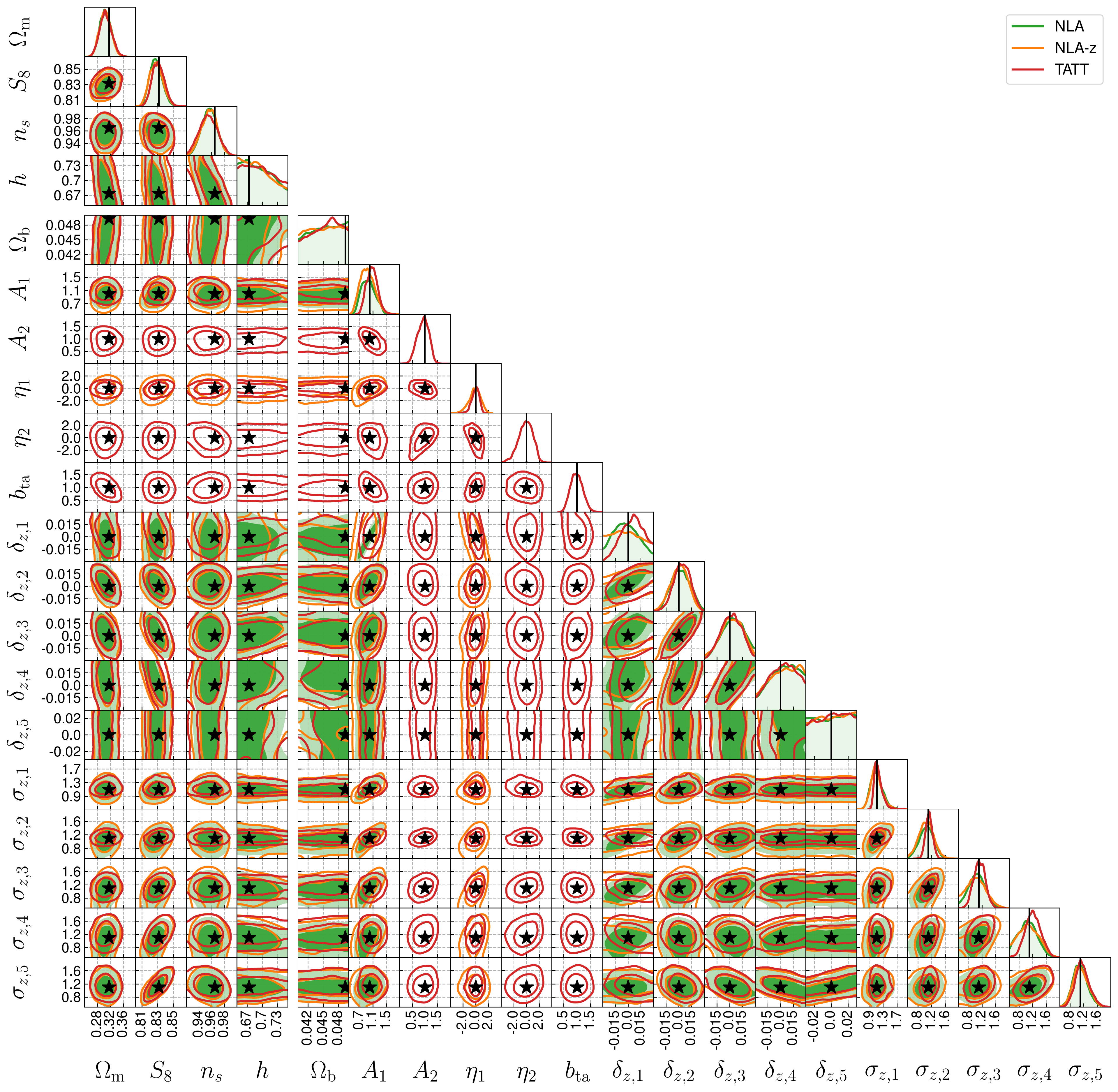}
    \caption{
    Posterior distribution (68\% and 95\% confidence level) with an error on $\sigma_z$ given by Equation \ref{eq:testsetup} for the Stage-IV setup.
    We use flat priors on cosmology, IA and all redshift parameters.
    }
    \label{fig:flat_prior}
\end{figure}

Here, we present the posterior distribution of the test setup presented in Section \ref{sec:include} when using flat priors on all redshift parameters.
The full chain is shown in Figure \ref{fig:flat_prior}.
The bias that is visible when having Gaussian priors on the redshift parameters dissappears when using flat priors.
We find that especially $\sigma_z$ can be constrained with standard deviations between 0.1 and 0.28.
For the lower redshift bins, we also get some constraints on the mean $\delta_z$ whereas the mean of the high redshift bins are barely constrained.
Although the bias is reduced, we find that the uncertainty on cosmological parameter such as $S_8$ increases considerably due to the wide priors in the redshift parameters.

\section{Impact of noise}\label{app:noise}
\begin{figure}
	\centering
    \begin{subfigure}[b]{0.48\textwidth}
         \centering
         \includegraphics[width=\textwidth]{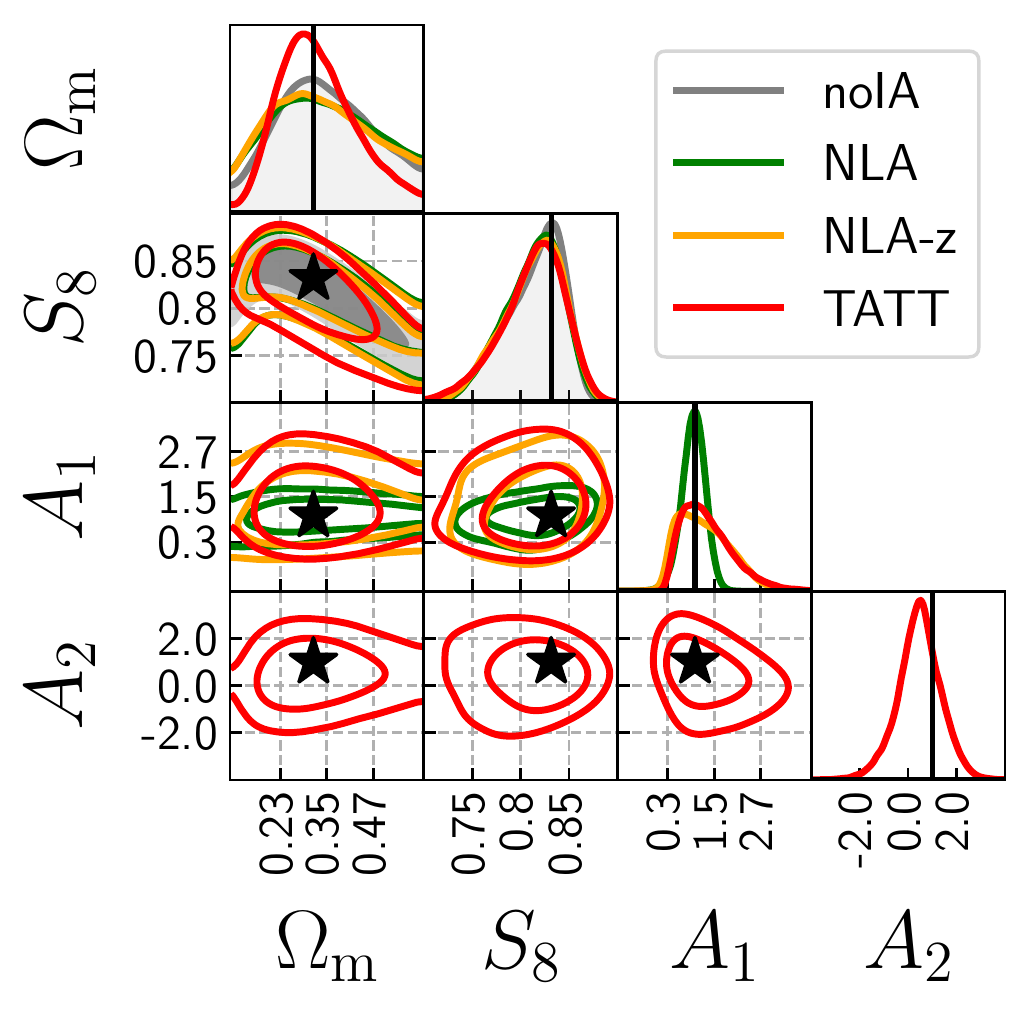}
         \caption{Stage-III}
         \label{fig:noise_DES}
    \end{subfigure}
    \begin{subfigure}[b]{0.48\textwidth}
         \centering
         \includegraphics[width=\textwidth]{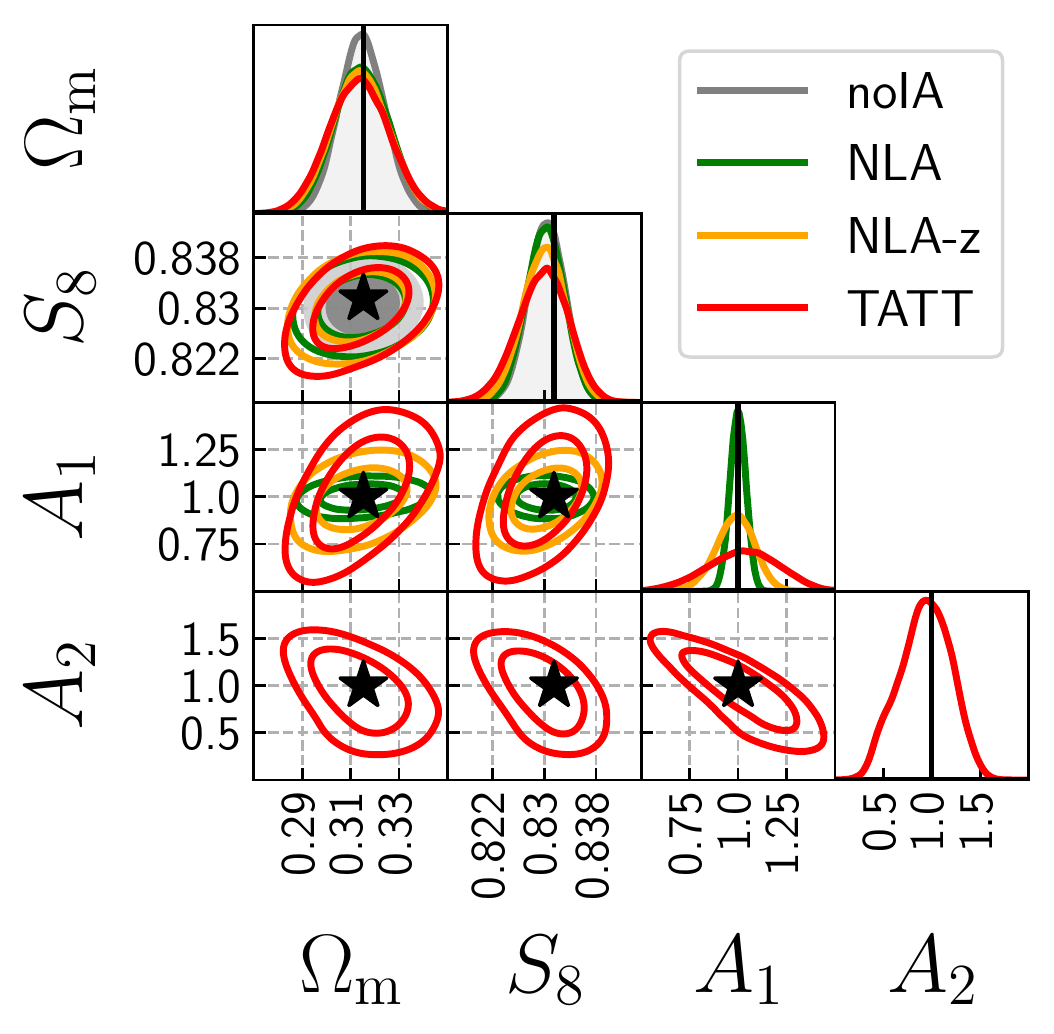}
         \caption{Stage-IV}
				\label{fig:noise_LSST}
    \end{subfigure}
	\caption{
	Posterior distribution (68\% and 95\% confidence level) of the stacked chains of 1000~different noise realisations with different IA models and with fixed redshift errors.
	The black star indicates the true value given in the mock observations.
	}
	\label{fig:noise}
\end{figure}

As described in Section~\ref{sec:req}, already noise can contribute to the bias.
The bias (as median of the best-fits) in terms of the uncertainty $\sigma_{\mathrm{av}}[S_8]$ is shown in Table~\ref{tab:noise}.
The stacked chains are shown in Figure~\ref{fig:noise}.
The bias of the noise is much higher for Stage-III than for Stage-IV.
The reason can already be seen when looking at the distribution of the stacked chains which is similarly shaped as the distribution of best-fits.
Although the distribution peaks close to the true value, the distribution is skewed towards lower $S_8$, i.e.\ in most cases, the $S_8$ that is found by a survey is lower than the truth.
For Stage-IV, the $S_8$ distribution is much more symmetric and almost not biased at all.
The noise bias we find in Stage-III surveys will be investigated in more detail in future work.
\begin{table}
	\centering
		\begin{tabular}{lllll}
		\toprule
		& \textbf{noIA} & \textbf{NLA} & \textbf{NLA-z} & \textbf{TATT} \\
		\midrule
		\textbf{Stage-III} & -0.17 & -0.34 & -0.40 & -0.47\\
		\textbf{Stage-IV} & -0.04 & -0.03 & -0.10 & -0.07\\
		\bottomrule
		\end{tabular}
	\caption{
	Bias $b_{\mathrm{n}}[S_8]/\sigma_{\mathrm{av}}[S_8]$ due to noise only in terms of the average uncertainty of a survey.
	}
	\label{tab:noise}
\end{table}

\section{Dependence of requirements on setup parameters}
\label{app:req_dep}
We present here how the requirements plot from Section \ref{sec:req} vary when certain parameters in the survey are changed. We investigate in particular the impact of different $\ell$ ranges and the impact of non-diagonal terms in the covariance matrix.

\begin{figure}
    \begin{subfigure}{\textwidth}
        \centering
	    \includegraphics[width=1\textwidth]{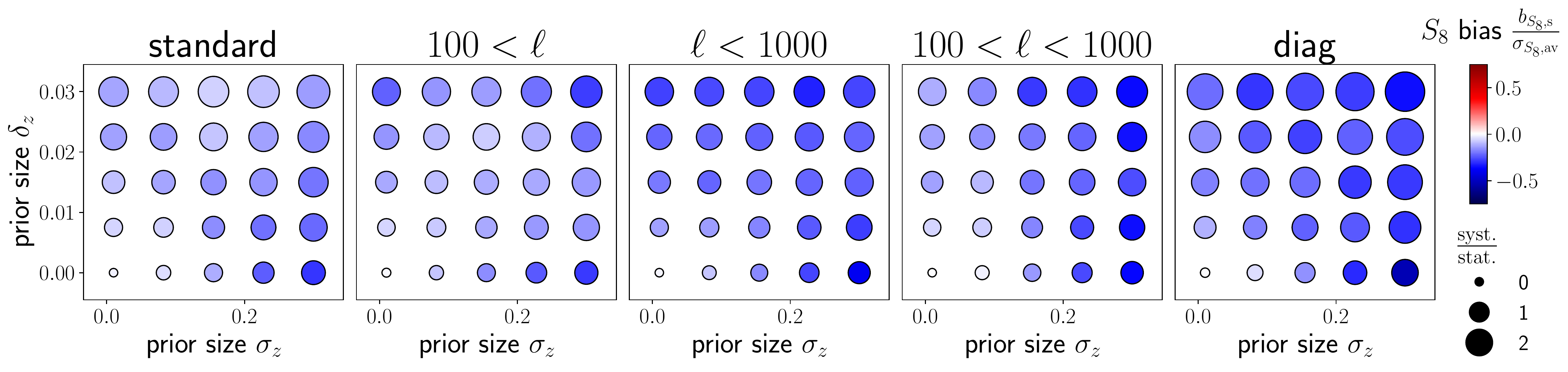}
        \caption{Ignoring $n(z)$ shape error in the model}
	    
    \end{subfigure}
    \begin{subfigure}{\textwidth}
        \centering
	    \includegraphics[width=1\textwidth]{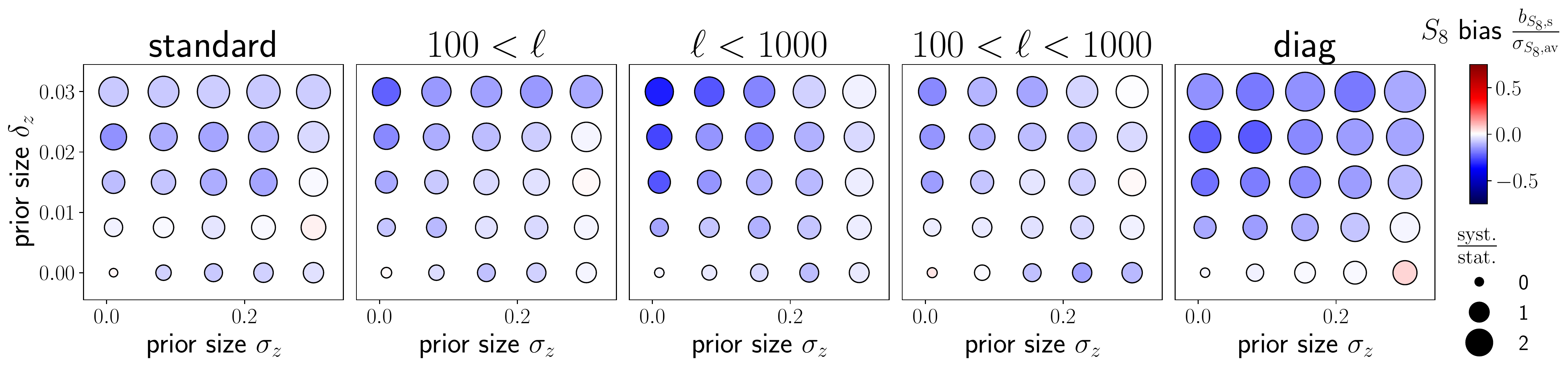}
        \caption{Including $n(z)$ shape error in the model}
	          
    \end{subfigure}
    \caption{
    Requirements for varying $\ell$-cuts and a diagonal covariance matrix using the NLA model.
    Note that the due wider prior width limits compared to Figure \ref{fig:req_LSST}, the colors and sizes are adjusted.
    }
    \label{fig:req_dep}
\end{figure}

The different setups are shown in Figure \ref{fig:req_dep}.
We compare our standard setups with $\ell$-cuts at 20 and 3000 with setups with different low and high cuts.
Further we compare our standard setup to a setup where we discard the non-diagonal terms of the covariance matrix in the analysis.
Generally, the requirements barely change for the different $\ell$-cuts and only slightly when discarding non-diagonal terms.
Note that this does not mean that the expected uncertainty is unchanged since we normalize by the average uncertainty of such a survey.
The uncertainty will increase when reducing the $\ell$-range.
But systematic and statistical uncertainty are affected by this in a very comparable way so that the ratio of the two does not change much.

We therefore expect our results to be quite stable regarding different $\ell$-ranges.
Nevertheless, we recommend to repeat such an analysis with the actual $\ell$-cuts of a specific survey in order to get a survey-specific requirement.

\end{document}